\documentclass[twocolumn, twocolappendix tighten, astrosymb, dvipsnames]{aastex631}
\usepackage{xspace}
\usepackage{color, soul}
\usepackage{graphicx}
\usepackage{amssymb}
\usepackage{amsmath}
\usepackage{float}
\usepackage{multirow}
\usepackage{textcomp}
\usepackage{gensymb}
\usepackage{hyperref}
\usepackage{natbib}
\usepackage{comment}
\usepackage{systeme}
\usepackage[Symbol]{upgreek}
\usepackage{xcolor}
\usepackage{orcidlink}
\usepackage{booktabs}

\newcommand{\chandra}{\textit{Chandra}\xspace}
\newcommand{\xmm}{\textit{XMM-Newton}\xspace}
\newcommand{\planck}{\textit{Planck}\xspace}

\received{January 12, 2025}
\revised{February 4, 2025}
\accepted{February 13, 2025}
\submitjournal{\apj}

\shorttitle{X-ray exploration of a low-mass cluster}
\shortauthors{Stroe et al.}

\graphicspath{{./}{figures/}}

\begin{document}

\title{PSZ2\,G181.06+48.47 I: X-ray exploration of a low-mass cluster with exceptionally-distant radio relics}

\correspondingauthor{Andra Stroe}
\email{andra.stroe@cfa.harvard.edu}

\author[0000-0001-8322-4162]{Andra Stroe}
\affiliation{Center for Astrophysics \text{\textbar} Harvard \& Smithsonian, 60 Garden St., Cambridge, MA 02138, USA}
\affiliation{Space Telescope Science Institute, 3700 San Martin Drive, Baltimore, MD 21218, USA}

\author[0000-0001-7509-2972]{Kamlesh Rajpurohit}
\affiliation{Center for Astrophysics \text{\textbar} Harvard \& Smithsonian, 60 Garden St., Cambridge, MA 02138, USA}

\author[0000-0001-8812-8284]{Zhenlin Zhu}
\affiliation{SRON Netherlands Institute for Space Research, Niels Bohrweg 4, 2333 CA Leiden, The Netherlands}
\affiliation{Center for Astrophysics \text{\textbar} Harvard \& Smithsonian, 60 Garden St., Cambridge, MA 02138, USA}
\affiliation{Leiden Observatory, Leiden University, Niels Bohrweg 2, 2300 RA Leiden, The Netherlands}

\author[0000-0002-3754-2415]{Lorenzo Lovisari}
\affiliation{INAF, Istituto di Astrofisica Spaziale e Fisica Cosmica, Via Alfonso Corti 12, 20133 Milano, Italy}
\affiliation{Center for Astrophysics \text{\textbar} Harvard \& Smithsonian, 60 Garden St., Cambridge, MA 02138, USA}

\author[0000-0002-9714-3862]{Aurora Simionescu}
\affiliation{SRON Netherlands Institute for Space Research, Niels Bohrweg 4, 2333 CA Leiden, The Netherlands}
\affiliation {Leiden Observatory, Leiden University, Niels Bohrweg 2, 2300 RA Leiden, The Netherlands}
\affiliation{Kavli Institute for the Physics and Mathematics of the Universe, The University of Tokyo, Kashiwa, Chiba 277-8583, Japan}

\author[0000-0002-5671-6900]{Ewan O'Sullivan}
\affiliation{Center for Astrophysics \text{\textbar} Harvard \& Smithsonian, 60 Garden St., Cambridge, MA 02138, USA}

\author[0000-0002-3984-4337]{Scott Randall}
\affiliation{Center for Astrophysics \text{\textbar} Harvard \& Smithsonian, 60 Garden St., Cambridge, MA 02138, USA}

\author[0000-0002-9478-1682]{William Forman}
\affiliation{Center for Astrophysics \text{\textbar} Harvard \& Smithsonian, 60 Garden St., Cambridge, MA 02138, USA}

\author[0000-0003-1949-7005]{Hiroki Akamatsu}
\affiliation{International Centre for Quantum-field Measurement Systems for Studies of the Universe and Particles (QUP), The High Energy Accelerator Research Organization (KEK), 1-1 Oho, Tsukuba, Ibaraki 305-0801, Japan}
\affiliation{SRON Netherlands Institute for Space Research, Niels Bohrweg 4, 2333 CA Leiden, The Netherlands}

\author[0000-0002-0587-1660]{Reinout van Weeren}
\affiliation{Leiden Observatory, Leiden University, PO Box 9513, 2300 RA Leiden, The Netherlands}

\author[0000-0002-5751-3697]{M. James Jee}
\affiliation{Department of Astronomy, Yonsei University, 50 Yonsei-ro, Seodaemun-gu, Seoul 03722, Republic of Korea}
\affiliation{Department of Physics and Astronomy, University of California, Davis, One Shields Avenue, Davis, CA 95616, USA}

\author[0000-0002-1566-5094]{Wonki Lee}
\affiliation{Department of Astronomy, Yonsei University, 50 Yonsei-ro, Seodaemun-gu, Seoul 03722, Republic of Korea}

\author[0000-0001-5966-5072]{Hyejeon Cho}
\affiliation{Department of Astronomy, Yonsei University, 50 Yonsei-ro, Seodaemun-gu, Seoul 03722, Republic of Korea}
\affiliation{Center for Galaxy Evolution Research, Yonsei University, 50 Yonsei-ro, Seodaemun-gu, Seoul 03722, Republic of Korea}

\author[0009-0009-4676-7868]{Eunmo Ahn}
\affiliation{Department of Astronomy, Yonsei University, 50 Yonsei-ro, Seodaemun-gu, Seoul 03722, Republic of Korea}

\author[0000-0002-4462-0709]{Kyle Finner}
\affiliation{IPAC, California Institute of Technology, 1200 E California Blvd., Pasadena, CA 91125, USA}

\author[0000-0003-2206-4243]{Christine Jones}
\affiliation{Center for Astrophysics \text{\textbar} Harvard \& Smithsonian, 60 Garden St., Cambridge, MA 02138, USA}

\begin{abstract}
Relics are diffuse, highly-polarized radio sources that trace merger-driven shocks at the periphery of merging galaxy clusters. The LOFAR survey recently discovered a rare example of double relics in the low-mass cluster PSZ2\,G181.06+48.47. Through a detailed exploration of new \chandra and \xmm observations, we reveal that PSZ2\,G181.06+48.47 has a lower mass ($M_{\rm 500,X}=2.32^{+0.29}_{-0.25}\times10^{14} \;{\rm M_{\odot}}$) than previously thought. Despite its cool global temperature of $kT_{500}=3.62^{+0.15}_{-0.07}$\,keV, PSZ2\,G181.06+48.47 is one of the most disturbed clusters in the \planck sample, with a complex morphological and thermodynamic structure. We discover a set of three discontinuities within $<500$\,kpc of the cluster center, and, from a surface brightness analysis, place $5\sigma$ upper limits of $\mathcal{M}_{NE}<1.43$ and $\mathcal{M}_{SW}<1.57$ for any shock associated with the relic locations. We also revise established scaling relations for double radio-relics by adding 12 new systems not included in previous work. The PSZ2\,G181.06+48.47 relics have the widest separation (scaled for $r_{500}$) of all known double-relic systems. The exceptional distance from the cluster center ($>r_{200}$), indicates the relics may be associated with shocks in the ``run-away" phase. We propose that this late-stage, post-apocenter merger is captured as the two subclusters with a mass ratio of 1.2--1.4 fall back into each other. The outer relic shocks were likely produced at the first core passage, while the inner discontinuities are associated with the second infall.
\end{abstract}


\section{Introduction}
\label{sec:intro}

Cluster mergers, a key mechanism for the growth of structure in the hierarchical formation process \citep[see][for a review]{Kravtsov2012}, are the most energetic events since the Big Bang, releasing huge amounts of energy, up to $\sim 10^{64}$\,ergs over a few Gyr timescales \citep{2001ApJ...561..621R, Hoeft2008}. Major cluster mergers heat the diffuse intracluster medium (ICM) and can potentially disrupt cooling flows \citep[e.g.][]{Gomez2002, Poole2006, Poole2008, ZuHone2011b, Valdarini2021}. 

Thermal and non-thermal components of the ICM such as shocks, cold fronts, turbulence, magnetic fields, and cosmic ray particles make clusters unique laboratories to study plasma physics \citep[e.g.][]{Sarazin1986, Sarazin2004, vanWeeren2019}. Thanks to the combination of sub-arcsecond angular resolution (0.5\arcsec, with \chandra) and sensitivity to diffuse extended thermal emission achievable by X-ray telescopes, the impact that merger events have on the plasma atmospheres and substructures can be directly mapped, for example, through sharp X-ray surface brightness edges, such as shocks and cold fronts \citep[e.g.][]{Markevitch2002, Markevitch2005, Markevitch2007, Owers2009, Ogrean2013b, Eckert2016, Akamatsu2017, XZhang2021, Diwanji2024}. Unlike shocks, where the gas density and temperature are higher in the downstream compared to the upstream, cold fronts are formed at the boundary of gas clouds moving through a hotter and rarefied surrounding medium \citep[see the review by][]{Markevitch2007}. Cold fronts have been detected in both merging and relaxed clusters \citep[e.g.,][]{Botteon2018, Ghizzardi2010, Markevitch2007, Paterno-Mahler2013}.

\begin{figure*}[!tbh]
    \centering
    \includegraphics[height=0.35\textwidth]{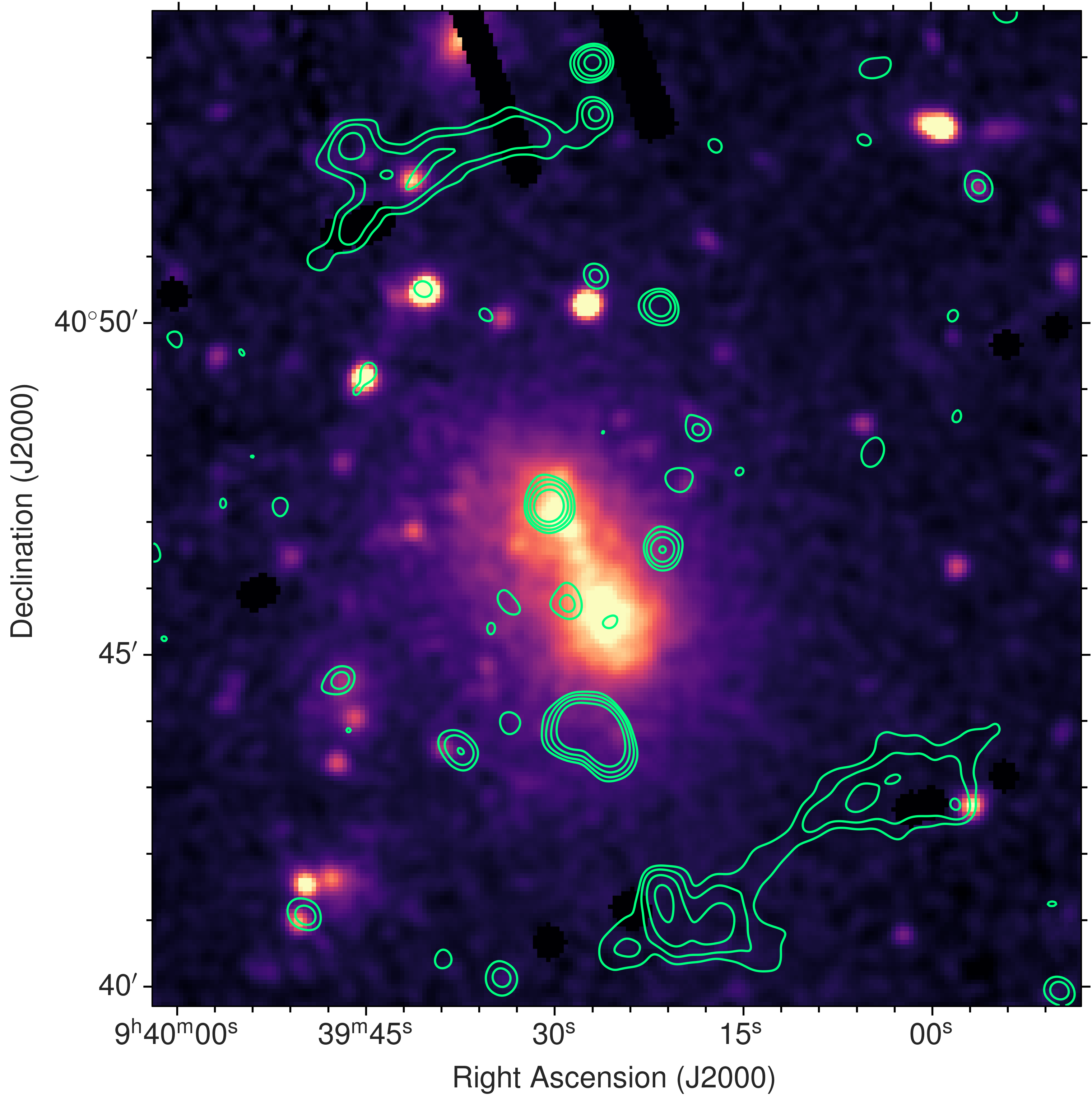}
    \includegraphics[height=0.35\textwidth]{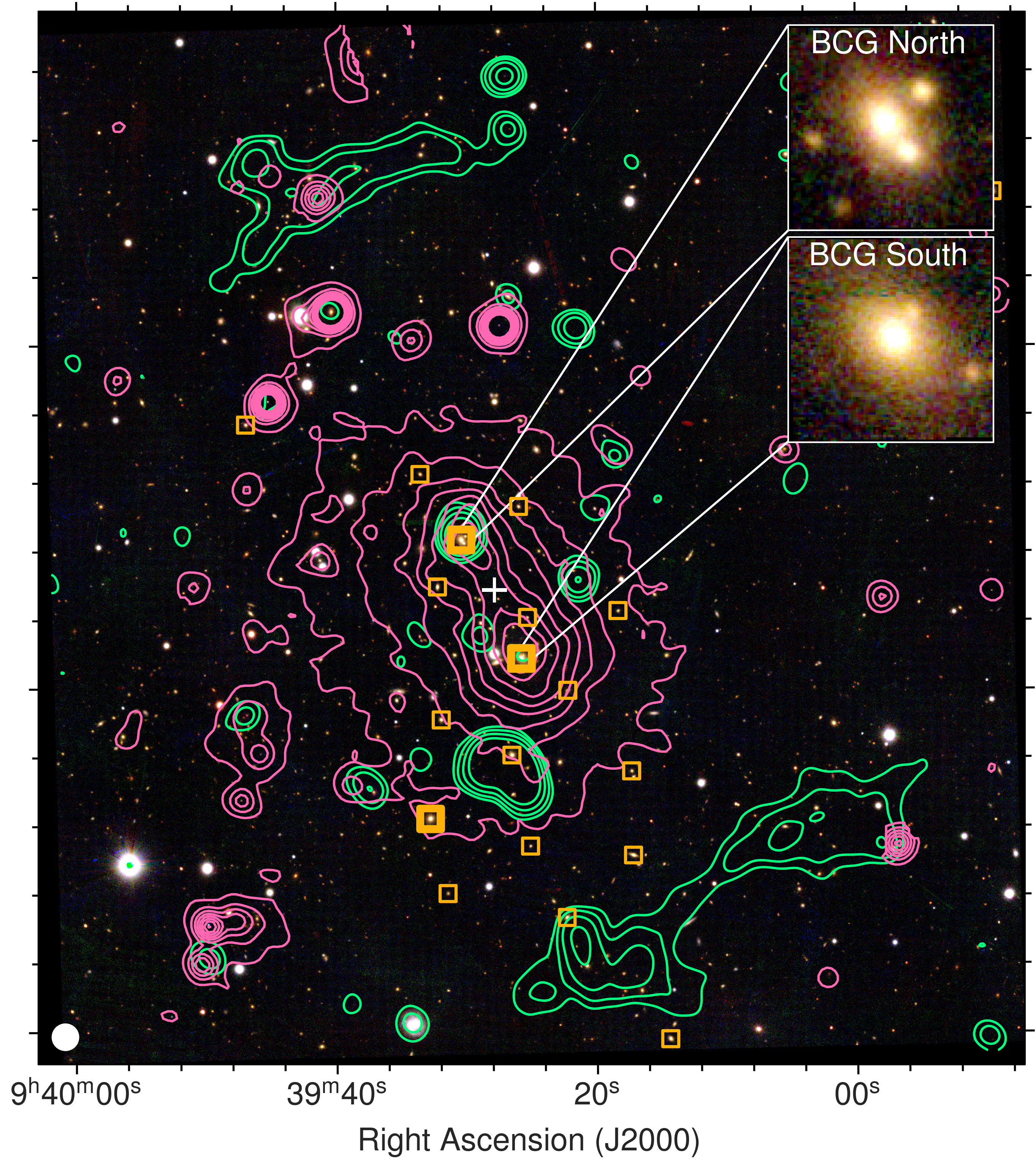}
    \includegraphics[height=0.35\textwidth]{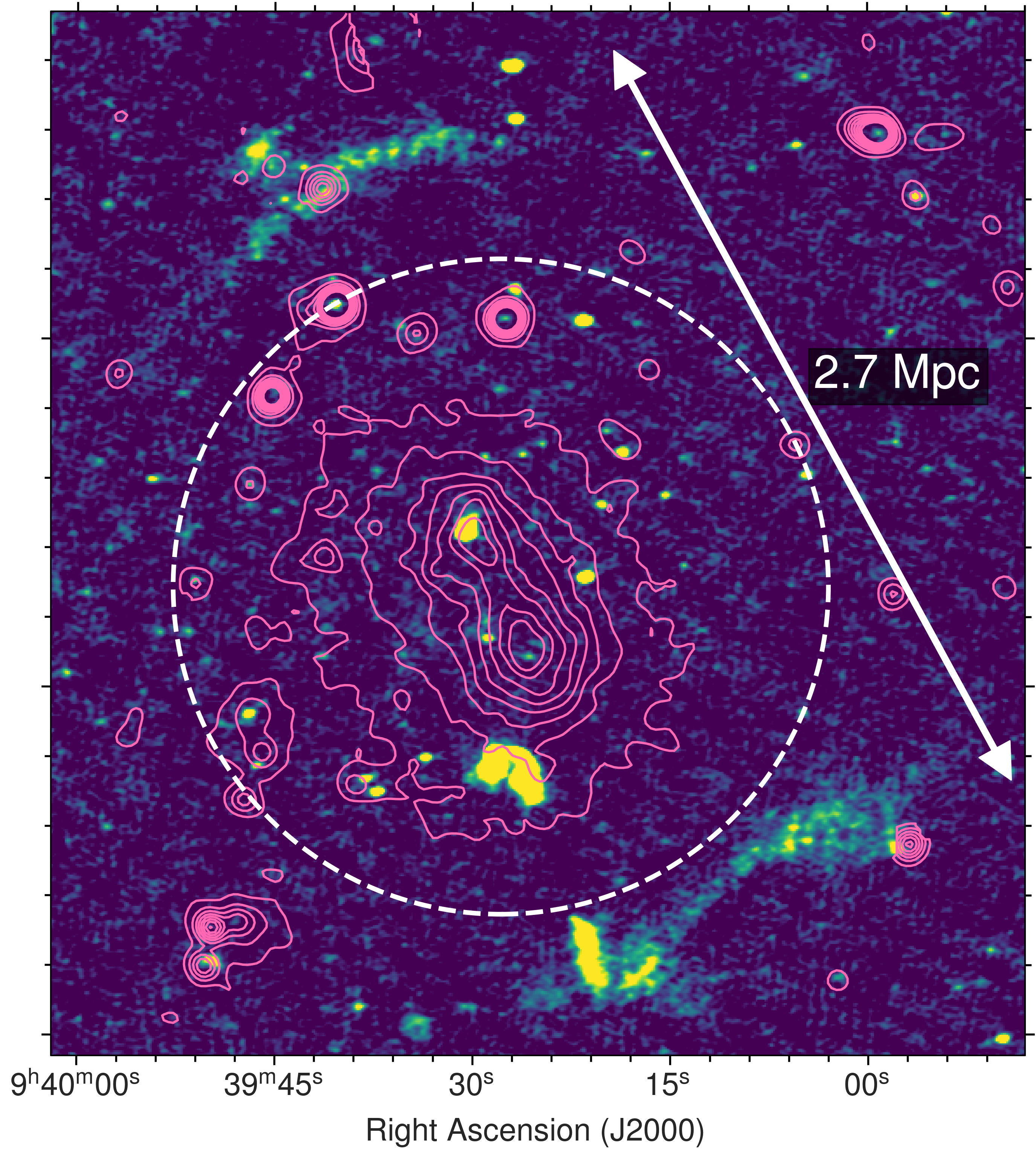}
    \caption{Multiwavelength view of PSZ2\,G181.06+48.47, showing the central $15\arcmin\times14\arcmin$ area. \textit{Left:} 0.5--2\,keV \xmm image of the cluster, with LOFAR 140\,MHz radio contours drawn at $[1, 2, 4, 8 ...]\times 3.0\sigma_{\rm rms}$, unveiling the cluster as a clear merger with an elongated, double-peaked X-ray morphology. \textit{Center:} RGB Pan-STARRS image, with radio (green) and \xmm (magenta, drawn at $[2, 4, 6, 8 ...]\times10^{-8}\,{\rm cts/s/pixel}$) contours overplotted. Saffron yellow squares mark spectroscopically-confirmed cluster members, including two BCGs roughly following the \xmm X-ray peaks. The white plus marks the center of the system. \textit{Right:} The 140\,MHz LOFAR image at 6{\arcsec} resolution reveals two spectacular diffuse radio sources on either side of the \xmm X-ray detection of the cluster ICM (contours in magenta). The white, dashed circle represents $r_{\rm 500,SZ}=1.06$\,Mpc.}
    \label{fig:multi}
\end{figure*}

As merger-induced shocks propagate towards the cluster outskirts during a cluster merger, a small fraction of the released energy is channeled into the (re)-acceleration of particles. In the presence of magnetic fields, these particles emit synchrotron radiation leading to the formation of Mpc-scale giant radio relics \citep{vanWeerenSci, Stroe2013, Rajpurohit2020a, Botteon2022a, Knowles2022, Jones2023, vanWeeren2019, Brunetti2014}. Located 0.5--2\,Mpc away from the cluster core, radio relics are typically characterized by a broad arc-like morphology complemented by a filamentary structure at high resolution \citep[e.g.][]{vanWeeren2019}. Relics are also highly polarized \citep[average polarization of 20--30\% at 1.4\,GHz;][]{Kierdorf2017, Stuardi2019}, which indicates highly-ordered magnetic field vectors. Simulations find that the fraction of $z=0-1$ clusters hosting relics increases rapidly from $\sim14\%$ for clusters with mass $M_{500}>{10}^{14}$\,M$_\sun$ to $\sim38\%$ with $M_{500}>5\times{10}^{14}$\,M$_\sun$ \citep{Lee2024}. As radio relics are preferentially found in the outskirts of massive merging clusters, it is challenging to detect them in low-density outer regions of the ICM \citep[e.g.][]{Botteon2022a, Duchesne2024}. Therefore, radio relics are a relatively rare occurrence \citep{Jones2023}.

In rare cases, merging systems can exhibit two relics on opposite sides of the cluster. To our knowledge, only 30 examples of merging clusters hosting double radio relics have been discovered so far \citep{deGasperin2014, vanWeeren2019, Botteon2020a, Jones2023}. For example, only about 1/3 of simulated radio relic clusters may host candidate double relics of relatively similar luminosity \citep[luminosity ratio $>0.3$,][]{Leeinprep}. More importantly, double relics are more likely to be observed in simple binary cluster mergers with a favorable orientation close to the plane-of-the-sky \citep{Golovich2019}, providing a unique laboratory to study the shocks with less interference by projection effects. 

While radio observations reveal the location of the relics, X-ray observations enable us to search for shock waves by identifying discontinuities in the surface brightness and thermodynamic profiles of the ICM. Following the first reported association between an X-ray shock and a radio relic \citep{Finoguenov2010}, dozens of radio relics have since been confirmed to be associated with X-ray shock fronts \citep[e.g.][]{Akamatsu2013, Akamatsu2015, Shimwell2015, Botteon2016a, Botteon2016b, vanWeeren2019, Ogrean2013a}. 

While many X-ray shocks associated with radio relics have been successfully studied, shocks created in low-mass clusters with lower ICM sound speeds and, hence, slower shock velocities remain under-explored \citep{Eckert2016, Akamatsu2017, Urdampilleta2018}. Such shocks probe a key region of parameter space and are critical to constraining shock acceleration models. For example, standard diffusive shock acceleration \citep[DSA, e.g.][]{Bell1978, Drury1983, Blandford1987}, where particles from the thermal pool are directly accelerated by the merger-induced shock front, has been proposed as an effective mechanism for the origin of radio relics \citep[for details see][]{Brunetti2014}. For low Mach number shocks such as those in clusters \citep[$\mathcal{M}<2$;][]{Ryu2019}, DSA is not sufficient to explain the observed bright emission associated with large radio relics due to the low acceleration efficiency \citep[e.g.][]{Botteon2020a}. Only a small fraction ($\sim$30\%) of known merging clusters hosting double relics have masses $\lesssim4\times10^{14}$\,M$_\odot$. To investigate the impact of low-velocity shocks on particle acceleration, case studies of relic-hosting low-mass clusters are crucial.

While the population of massive clusters has been extensively characterized in the X-rays, multiwavelength datasets are increasingly detecting low-mass ($\leq 4 \times 10^{14} M_{\odot}$) clusters. In this paper, we focus on a detailed investigation of a low-mass cluster PSZ2\,G181.06+48.47 in order to uncover its dynamical state and evolution history (Figure~\ref{fig:multi}). The cluster has been little studied in the X-ray band. Originally discovered and confirmed through optical photometric methods at $z\sim0.24$ \citep[e.g.][]{Koester2007, Rykoff2014}, the cluster was detected by \planck and presented in the catalog of Sunyaev-Zel'dovich sources \citep{Planck2016} as a low-mass cluster ($M_\mathrm{SZ}=(4.2\pm0.5)\times10^{14}\,\mathrm{M_{\odot}}$; $r_{\rm 500,SZ}\footnote{The radius within which the mean density is 500 times the critical density at the cluster redshift.}=(1.06\pm0.05)\,\rm{Mpc}$). Double, giant (1.2--1.3\,Mpc) radio relics on either side of PSZ2\,G181.06+48.47 and a candidate halo at the center were discovered recently \citep{PSZ_radio}. The relics were first reported in the LOw Frequency ARray (LOFAR) Two-meter Sky Survey data for 309 \planck SZ-selected galaxy clusters \citep[LoTSS-DR2;][]{Botteon2022a}. Both relics are located at very large distances, over $1.2$\,Mpc from the cluster center. The cluster, coinciding with the location of diffuse radio sources, appears to be a merging system. However, with a \planck detection at a signal-to-noise (S/N) of $\sim5$, the detailed merger scenario of PSZ2\,G181.06+48.47 remained unknown. Nevertheless, the low cluster mass reported by \planck, which implies low particle densities, presents challenges in explaining the positions of relic candidates far apart from each other \citep{Brunetti2014, Kale2017}. PSZ2\,G181.06+48.47 thus presents a unique opportunity to study particle acceleration in the low-mass, low-density regime.

We aim to paint a comprehensive picture of PSZ2\,G181.06+48.47 through multiwavelength observations spanning the electromagnetic spectrum. In this first paper of the series, we utilize \chandra and \xmm to unambiguously confirm PSZ2\,G181.06+48.47 as a merging cluster, measure its detailed properties and characterize its substructure. We search for ICM discontinuities to understand the cluster dynamics. By accurately determining the cluster mass distribution, we also explore viable merger scenarios. \citet{PSZ_radio} presents new multi-band radio observations of the cluster to determine the properties of the diffuse radio sources and derive key insights into the underlying particle acceleration mechanism. \citet{PSZ_WL} leverage a weak-lensing (WL) analysis, complemented by tailored hydrodynamical simulations, to obtain accurate cluster mass estimates, as well as to constrain the merger scenario.

The structure of the paper is as follows. In Section~\ref{sec:data}, we outline the observations and data reduction. The X-ray imaging and spectral analyses and methods are presented in Section~\ref{sec:analysis}. We discuss the system and subcluster properties, including the estimation of the cluster mass, the subcluster mass ratios, and the morphologic and thermodynamic structures in Section~\ref{sec:clusterproperties}. We analyze our newly discovered discontinuities in Section~\ref{sec:discontinuities}. In Section~\ref{sec:comparison}, the cluster double relics are placed in the context of other known double relics. We discuss the merger scenario for PSZ2\,G181.06+48.47 in Section~\ref{sec:mergerscenario}. Finally, we summarize our findings in Section~\ref{sec:conclusion}. We adopt a $\Lambda$CDM cosmology with $\Omega_m=0.286$ and $H_{0} = 69.6\,\mathrm{km\,s^{-1}\,Mpc^{-1}}$, resulting in a physical scale $1\arcmin = 225\,\mathrm{kpc}$ at the cluster redshift $z=0.234$ \citep{Wright2006}. All errors are quoted at the 1$\sigma$ level unless otherwise noted. We use a system center defined to lie between the two subclusters at ${\rm RA}=144.866\degree$ and ${\rm Dec}=40.774\degree$.

\begin{figure*}[!tbh]
    \centering
    \includegraphics[width=\textwidth]{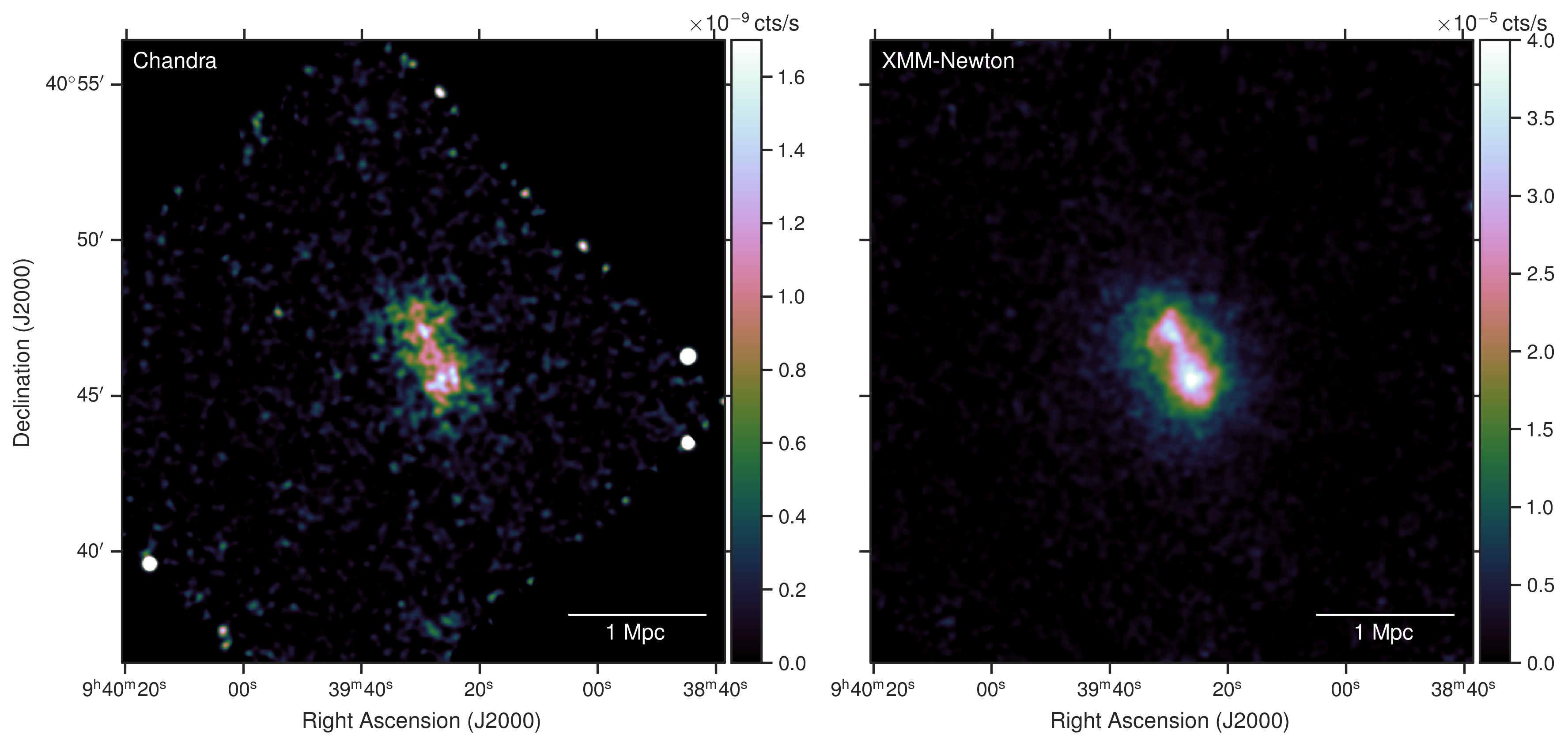}
    \caption{X-ray exposure corrected and background subtracted flux maps of PSZ2\,G181.06+48.47 in the 0.5--2\,keV band from \chandra (\textit{Left}) and 0.7--2\,keV band from \xmm (\textit{Right}). Point sources were refilled for visualization purposes. Both maps were smoothed with a Gaussian kernel of $15''$ FWHM and shown in a linear color scale.}
    \label{fig:xray_image}
\end{figure*}

\section{Observations and Data reduction}
\label{sec:data}

\subsection{\xmm}
\label{sec:xmm}

PSZ2\,G181.06+48.47 was observed for 120\,ks by the \xmm European Photon Imaging Camera (EPIC) MOS+pn in May 2020 (PI: A. Stroe, ObsID: 0862970101 and 0862970201; see Table~\ref{tab:xray_obs}). We reduced the data using the XMM-Newton Science Analysis System (SAS) v20.0.0. MOS and pn event files were obtained from the observation data files with the tasks \texttt{emchain} and \texttt{epchain}. The out-of-time event file of pn was also produced by \texttt{epchain}.

The data were cleaned for periods of high background using the tasks \texttt{mos-filter} and \texttt{pn-filter}. We did not find any significant time intervals affected by flares, such that after this cleaning, the net exposure times of MOS1, MOS2, and pn were $\sim$111\,ks, $\sim$110\,ks, and 92\,ks, respectively. Following standard filtering techniques, we kept the single, double, triple, and quadruple events in the MOS data (pattern$\leq$12) and restricted the pn data to single and double events only (pattern $\leq4$). Additionally, we also excluded all the CCDs in the so-called anomalous state \citep[for more details, see][]{Kuntz_Snowden_2008}. Point-like sources were detected using the task \texttt{edetect\_chain} and excluded from the event files (see Figure~\ref{fig:extract_regions}). We used the Filter Wheel Closed (FWC) data, properly rescaled to match our observation (see Section~\ref{sec:xray_spec}), as instrumental backgrounds event files. 

The background-subtracted and vignetting-corrected X-ray image in the 0.7--2.0\,keV band shown in Figure~\ref{fig:xray_image} was obtained using a binning of 40 physical pixels (corresponding to a resolution of 2\arcsec). For visualization purposes, we refilled the point-source regions using the CIAO task \texttt{dmfilth}.

\begin{deluxetable*}{c c c c c c c}[!thbp]
\tablecaption{X-ray observational details.\label{tab:xray_obs}}
\tablehead{
Telescope & ObsID & Instrument & \multicolumn{2}{c}{Pointing Coordinates} & Start Date & Valid Exp. \\
& & & R.A. & Decl. & & \\
& & & (deg)      & (deg)      & & (ks) }
\startdata
\multirow{6}{*}{\xmm}  & \multirow{3}{*}{0862970101} &  EPIC-MOS1 &  \multirow{3}{*}{144.86321} &  \multirow{3}{*}{40.75850} & \multirow{3}{*}{2020 May 5} & 41 \\
&  &  EPIC-MOS2 & & & & 40   \\
&  &  EPIC-PN   & & & & 34 \\\cline{2-6}
&  \multirow{3}{*}{0862970201} & EPIC-MOS1 & \multirow{3}{*}{144.86375} & \multirow{3}{*}{40.75825} & \multirow{3}{*}{2020 May 6} & 70 \\
&  &  EPIC-MOS2 & & & & 69  \\
&  &  EPIC-PN   & & & & 58 \\\hline
\chandra & 22650 & ACIS-I  & 144.88675 & 40.7705 & 2019 Sep 23 &  54
\enddata
\end{deluxetable*}

\subsection{\chandra} 
\label{sec:chandra}

PSZ2\,G181.06+48.47 was observed by \chandra ACIS-I for $\sim$60\,ks in September 2019 (PI: A. Stroe, ObsID: 22650). All data were reprocessed using the CIAO data analysis package, version 4.13, and the calibration database (CALDB 4.9.4) distributed by the Chandra X-ray Observatory Center. We removed  cosmic rays and bad pixels from all level-1 event files using the \texttt{chandra\_repro} CIAO tool with VFAINT mode background event filtering. We extracted the light curve in the 9--12\,keV band for each observation to examine possible contamination from flare events. To remove the time intervals with anomalous background, we applied the CIAO tool \texttt{deflare} to filter out the times where the background rates exceed $\pm 2 \sigma$  from the mean. We obtained a 54\,ks clean exposure as a result.

We utilized stowed background files to estimate the non-X-ray background (NXB). All stowed background event files were combined and reprocessed with \texttt{acis\_process\_events} using the updated gain calibration files. For the whole field of view (FOV), we scaled the NXB to match the 9--12\,keV count rate of the observation.

To search for surface brightness edges, we created \chandra images in the 0.5--2\,keV band. Background images were generated from the stowed observation files. We calculated the exposure maps using a weighted spectrum file generated by \texttt{make\_instmap\_weighted}, where the spectral model is an absorbed \texttt{apec} model with $kT = 5$\,keV. 

The smaller point spread function (PSF) of \chandra enables us to cleanly separate point-like and diffuse sources. We identified two extended diffuse clumps located north and south of the cluster, at distances beyond $r_{\rm 500,SZ}$. Since these extended sources are located along the merger axis, we examined their potential association with the larger-scale environment of PSZ2\,G181.06+48.47. Both the northern and southern clumps coincide with background clusters at $z\sim0.49$ \citep{Wen2015} and $z\sim0.32$ \citep{Zou2021, PSZ_WL}, respectively. Similar clump-like diffuse features observed, for example, in the outskirts of A\,133 and A\,1795 are predominantly attributed to background clusters that are projected along the line of sight \citep[LOS;][]{Zhu2023, Kovacs2023}. Both clumps were also excluded from any further analysis.

We applied the \texttt{wavdetect} algorithm on the \chandra counts image in the 0.5--2\,keV band, adopting a false-positive probability threshold of $10^{-6}$ and wavelet scales of 1, 2, 4, 8, 16, 32 pixels. We excluded the extended clumps as well as the point sources detected above. We manually adjusted the \texttt{wavdetect} regions for bright sources to safely exclude the surface brightness wings of their emission. The \chandra flux map, shown in the left panel of Figure~\ref{fig:xray_image}, was obtained after correcting the NXB-subtracted count image with the combined exposure map and filling the point sources with \texttt{dmfilth}. 

\subsection{Ancillary observations}

We use multiwavelength data to aid in the interpretation of the X-ray observations, including 144\,MHz images from the LOFAR Two-metre Sky Survey \citep[LoTSS;][]{Shimwell2019, Botteon2022a}, optical \textit{g}, \textit{r}, and \textit{i} imaging from the Panoramic Survey Telescope and Rapid Response System survey \citep[Pan-STARRS;][]{2020ApJS..251....7F}, and optical spectroscopy from the Sloan Digital Sky Survey Data Release 18 \citep[SDSS DR 18;][]{2023ApJS..267...44A}. We refer the reader to \citet{PSZ_radio} and \citet{PSZ_WL}, respectively, for details on the new multi-band radio and optical observations, respectively on the cluster.

\begin{figure*}[!tbh]
    \centering
    \includegraphics[width=\textwidth]{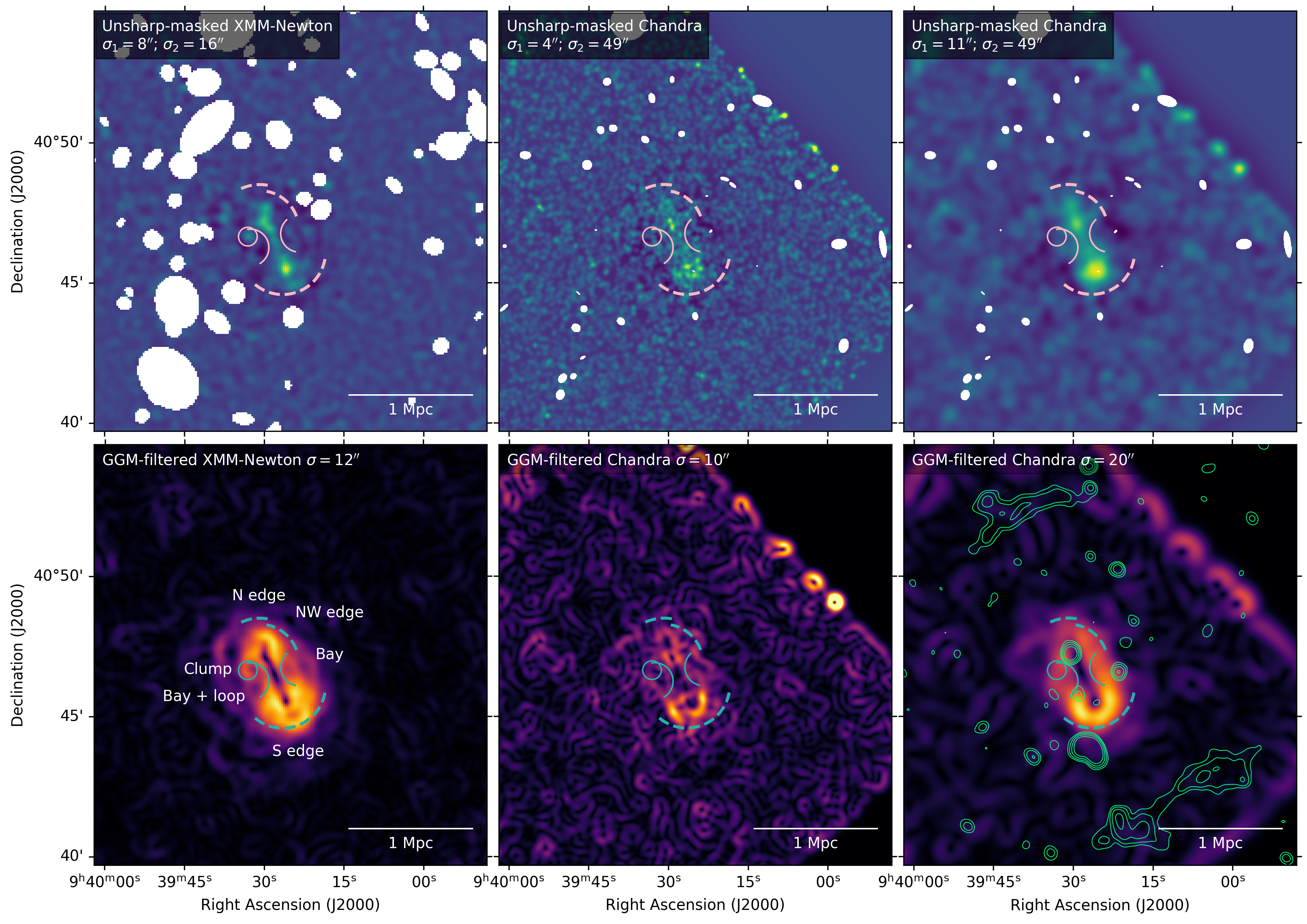}
    \caption{\textit{Top row:} Unsharp-masked \xmm and \chandra maps, obtained by subtracting images convolved with Gaussians of different widths. The point sources excluded for this analysis are denoted with white solid ellipses. \textit{Bottom row:} GGM filtered \xmm and \chandra images using a range of widths. LOFAR 144\,MHz contours are shown in the bottom-right panel. The unsharp-masked and GGM-filtered images highlight the presence of X-ray discontinuities roughly following the NE-SW merger axis of the cluster.}
    \label{fig:um_ggm}
\end{figure*}

\section{Analysis and methods}
\label{sec:analysis}

\subsection{Unsharp mask and Gaussian gradient magnitude filtered images}

To search for smaller-scale substructures and discontinuities, we first employ the unsharp masking and Gaussian Gradient Magnitude (GGM) filtering techniques \citep{Fabian2003, Sanders2016}. An unsharp masked image is typically produced by subtracting the image smoothed with a large Gaussian kernel from the same image convolved with a smaller-width Gaussian. We explored a range of smoothing widths and present the unsharp-masked maps with the most distinguishable features in the upper panel of Figure~\ref{fig:um_ggm}. The GGM filter computes the gradient magnitude of an image convolved with a Gaussian of a particular width. We show the GGM-filtered \xmm (convolved with a 12{\arcsec} kernel) and \chandra (convolved with a 10{\arcsec} and 20{\arcsec} kernel) images in the lower panels of Figure~\ref{fig:um_ggm}. To avoid the detection of spurious discontinuities, we cosmetically ``heal" point sources by using a 5-pixel disk footprint and filling previously-masked point sources with pixel values randomly selected from the surrounding region \citep{Sanders2016b}.

\subsection{Surface brightness profiles}

\begin{figure}[!tbh]
\includegraphics[width=\linewidth]{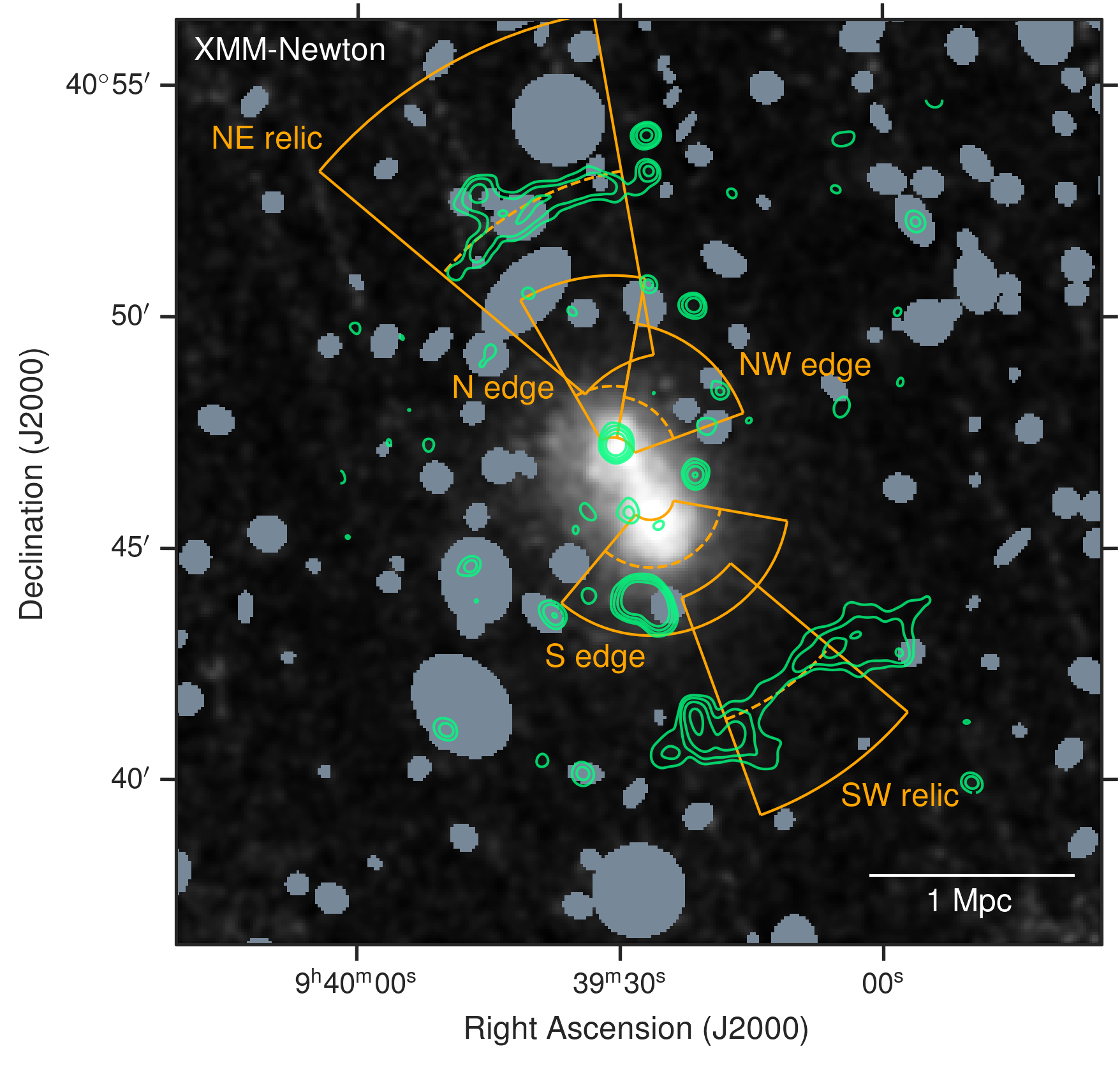}
\caption{\xmm flux image with LOFAR 140\,MHz green contours overplotted. The regions used to extract the surface brightness profiles across the central discontinuities and relics are presented as orange sectors, with the overlaid dashed lines highlighting the radius of identified breaks. The areas masked for the surface brightness analysis are filled in with slate gray.}
\label{fig:extract_regions}
\end{figure}

To detect edges and characterize their underlying density jumps, we extract surface brightness profiles from the point-source masked \xmm 0.5--2.0\,keV band image. We investigate surface brightness profiles extracted from a set of $30^\circ$-wide sectors covering the entire radial extent of the PSZ2\,G181.06+48.47 system. We use this blind approach to find discontinuities and refine the sectors used to extract the corresponding surface brightness profiles. We also investigate the presence of any discontinuities at the location of the radio relics. The refined sectors used for the extraction are shown in Figure~\ref{fig:extract_regions}.

The extraction and modeling of the surface brightness profiles were conducted using the \texttt{pyproffit} (version 0.8) package by \citet{Eckert2020}. For each surface brightness profile extracted, we explored a range of binning widths, start and stop radial distances, and start and stop angles to ensure the robustness of the strength and location of any jumps identified. We set the bin width at $7''$ to Nyquist sample the \xmm PSF.

We modeled the \xmm PSF as a King function by setting \texttt{prof.PSF(psffunc=xmm\_fking)}:
\begin{equation}
{\tt xmm\_fking} = \left(1+\left(\frac{x}{r_0}\right)^2\right)^{-\alpha},
\end{equation}
where $r_0 = 0.0883981\arcmin$ is the core radius and $\alpha=1.58918$ is the outer slope. 

To search for edges, the profiles were fit using $\chi^2$ minimization, with both a broken power-law, 3D density model converted to emissivity by projecting along the LOS (\texttt{pyproffit.BknPow}) and a single power law model (\texttt{pyproffit.PowerLaw}). The broken power-law model is defined as:
\begin{align}
   I(r) &= I_0 \int F(\omega)^2 d\ell + B, \\
   F(\omega) &= \begin{cases}
    \omega^{-\alpha_{1}}, &  \omega < r_{\rm f} \\ 
    \frac{1}{C} \omega^{-\alpha_{2}}, & \omega \geq r_{\rm f}
\end{cases}
\end{align}
where $r$ represents the projected radius, $l$ denotes the distance along the LOS and $\omega$ is defined by the equation $\omega^2 = r^2 + l^2$. The $r_{\rm f}$ indicates the radial position of the discontinuity, while $B$ represents the sky background. $\alpha_1$ and $\alpha_2$ are the slopes of the power law in the downstream and upstream regions, respectively. The $C=n_{\rm e,post}/n_{\rm e,pre}$ parameter represents the electron density jump between the post- ($n_{\rm e,post}$) and pre-discontinuity ($n_{\rm e,pre}$) regions. 

The single power law is defined as:
\begin{equation}
I(r) = I_0 \left(\frac{r}{r_{\rm p}} \right)^{-\alpha} + B,
\end{equation}
where $\alpha$ is the slope of the power law and $r_{\rm p}$ is the pivot point.

When fitting models to surface brightness profiles within the central areas of the system (within $6\arcmin$ of the center), all parameters were allowed to vary, except for the background $B$ which was fixed to 0, since the scaled background is subtracted by \texttt{pyproffit} prior to fitting. For the radio relics, the position of the discontinuity $r_{\rm f}$ is fixed to the outer edge of the radio arcs, at the putative location of the shocks. We also constrained the jump to values above 1, to avoid nonphysical fits where the density increases with radius.

Following the same regions used for the surface brightness profiles (Figure~\ref{fig:extract_regions}, with edges denoted by dashed lines), we extracted spectra from regions inside and outside of the edges to measure the ICM temperatures. See Section~\ref{sec:xray_spec} for details on the temperature modeling.

\subsection{X-ray spectral analysis}
\label{sec:xray_spec}

We study the thermodynamic structure of PSZ2\,G181.06+48.47 through a spectroscopic analysis of the \xmm data. Spectral modeling was carried out using \texttt{XSPEC} \citep[v12.14.1,][]{Arnaud1996}. The spectra were grouped to have at least one count per channel to use C-statistics (i.e., modified Cash statistics; \citealt{Cash1979}) as implemented in \texttt{XSPEC}. The ICM was modeled as an absorbed thermal plasma in collisional ionization equilibrium using the \texttt{phabs*apec} model with metallicities from \cite{2009ARA&A..47..481A} and total column density of $N_{\rm H} = 1.39 \times \mathrm{10^{20}\,cm^{-2}}$ \citep{Willingale2013}\footnote{Online calculator for Galactic column density: \url{https://www.swift.ac.uk/analysis/nhtot/index.php}}. The \texttt{apec} model is based on AtomDB v3.0.9 \citep{Foster2012}.

The modeling of the background was done following \citet{2019MNRAS.483..540L} with the changes highlighted in \citet{Lovisari2024}. We provide a brief summary of the procedure here and refer the interested reader to \citet{2019MNRAS.483..540L} and \citet{Lovisari2024} for a detailed description. The cosmic X-ray background was modeled by fitting simultaneously the \xmm spectra with the \textit{ROSAT} All-Sky Survey spectra extracted, using the \texttt{sxrbg}\footnote{\url{https://heasarc.gsfc.nasa.gov/cgi-bin/Tools/xraybg/xraybg.pl}} tool \citep{Sabol2019}, from a region beyond the estimated virial radius of the cluster. The non-vignetted quiescent particle background was estimated using the filter wheel closed observations after renormalizing them to match our observations following the strategy presented in \citet{2009ApJ...699.1178Z}. Finally, we included in the background model an extra broken power-law (folded only with the Redistribution Matrix File, RMF), to account for the residual soft proton contamination affecting \xmm even after filtering the flare events.

\subsection{Measuring global and subcluster properties}
\label{sec:propmethods}

To obtain the global temperature, we extracted \xmm spectra from the region within $r_{\rm 500,SZ}$ of PSZ2\,G181.06+48.47 and fit them in the 0.5--12\,keV and 0.5--14\,keV energy range for MOS and pn, respectively. Using the spectral fit information, the global luminosity was also computed using \texttt{XSPEC}.

To obtain the temperatures of the two subclusters, we extracted the spectra within a circular region around each of the two X-ray peaks. We define the main peak to lie within the southern subcluster at ${\rm RA}=144.85863\degree$ and ${\rm Dec}=40.75768\degree$, while the second peak lies within the northern subcluster and is located at ${\rm RA}=144.87223\degree$ and ${\rm Dec}=40.78241\degree$. We tested radii of $30''$, $45''$, and $60''$ within the two X-ray peaks, noting that, at $60''$, the areas for the two subclusters start overlapping.

\subsection{Thermodynamic maps}

To investigate the temperature, electron density, pressure, and entropy distribution of the merging cores, we follow the method presented in \citet{Lovisari2024}, with the regions obtained using the Weighted Voronoi Tessellation (WVT) binning algorithm by \citet{2006MNRAS.368..497D}, which is a generalization of the \citet{2003MNRAS.342..345C} Voronoi binning algorithm, by requiring a signal-to-noise S/N$\sim$30. While there are uncertainties associated with any choice, in deriving the density map, we assume that each emitting region has homogeneous properties and a volume of $V=2 A \sqrt{r_{500,P}^2-X^2-Y^2}$, where $A$ is the area of the region, and $X$ and $Y$ are the projected distances in the east-west and north-south directions from the cluster center, respectively. 

\begin{figure*}[!tbh]
    \centering
    \includegraphics[width=0.49\textwidth]{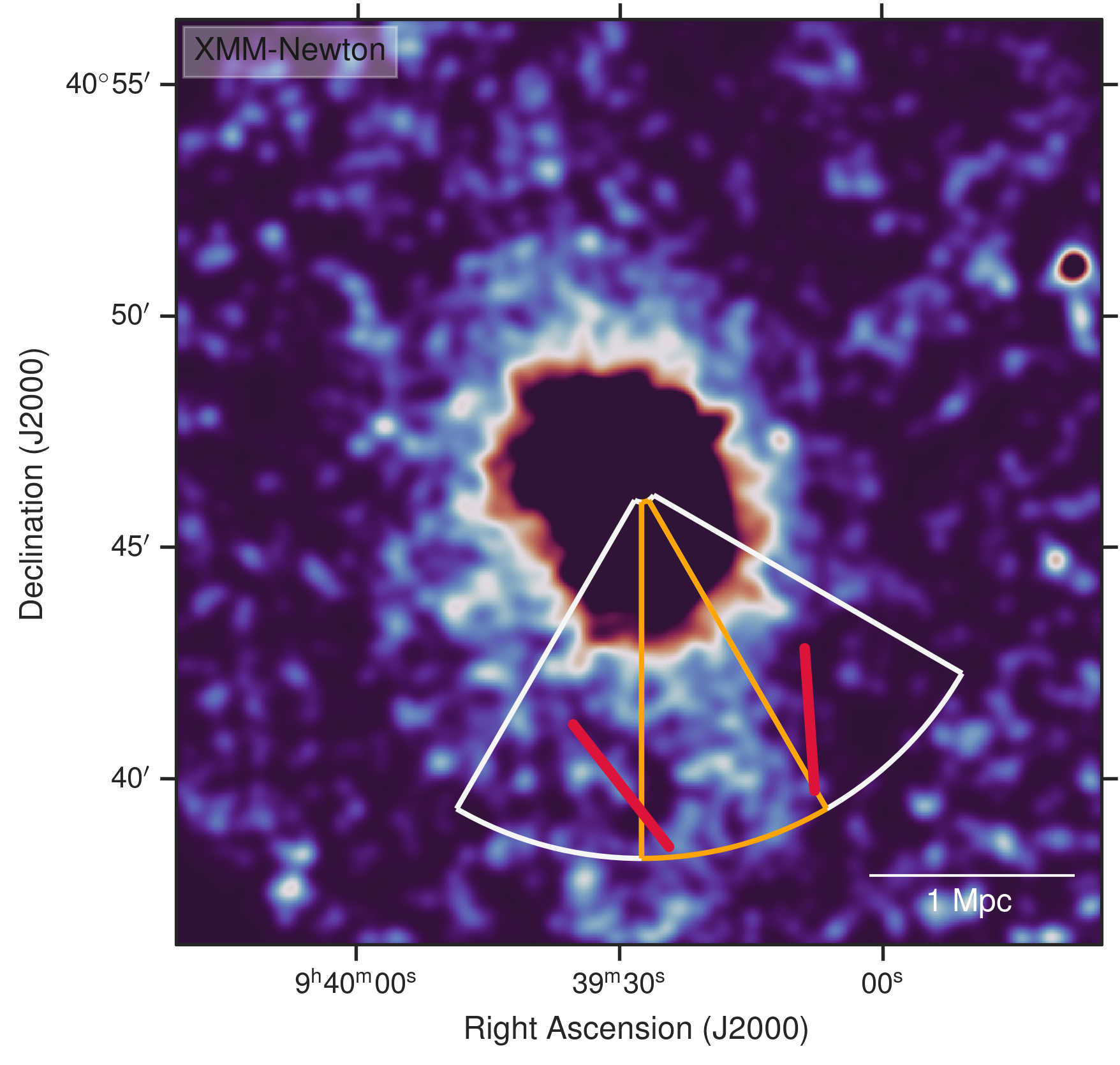}
    \includegraphics[width=0.49\textwidth]{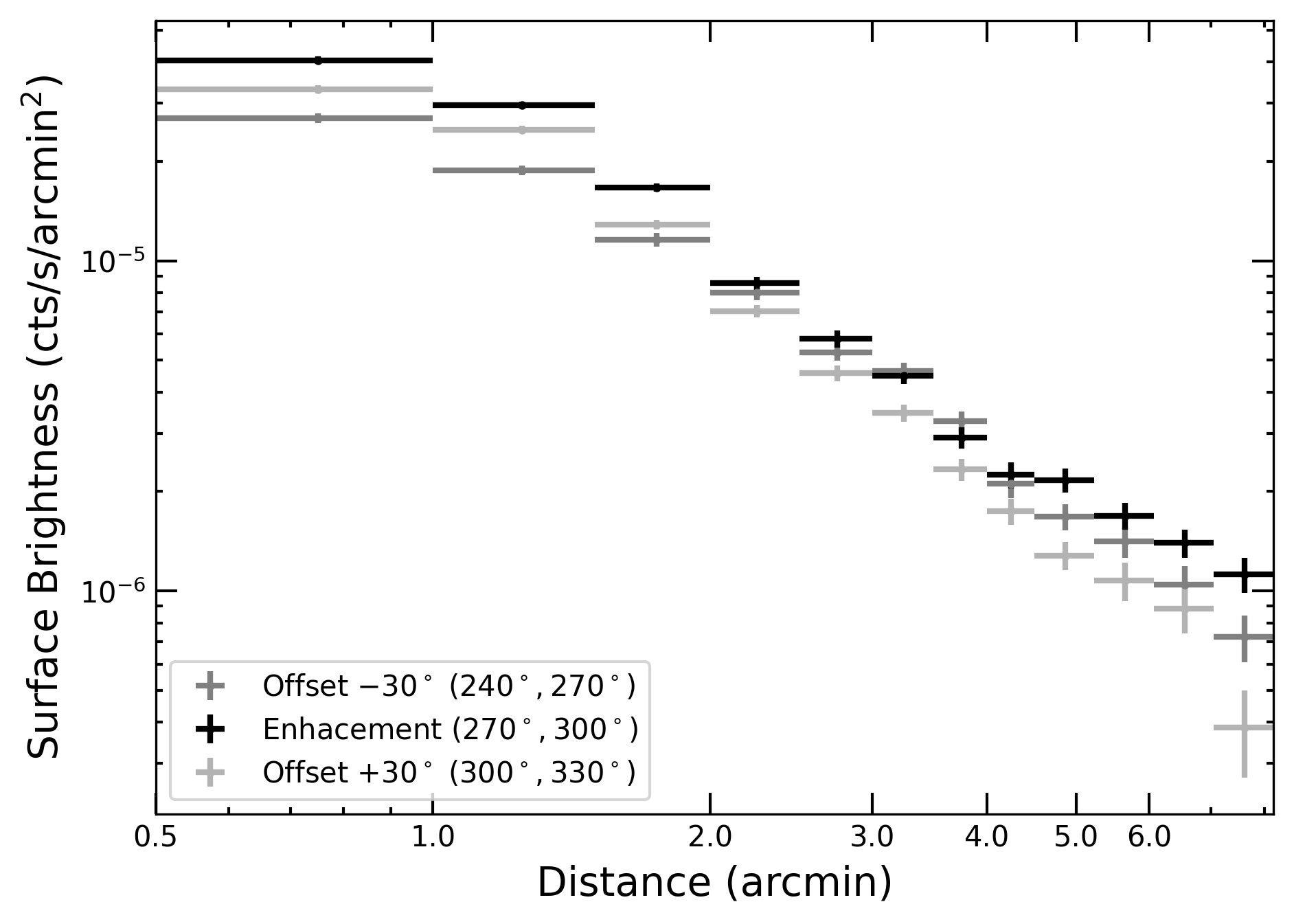}
    \caption{Surface brightness ``trail" towards the south of the PSZ2\,G181.06+48.47 system. \textit{Left panel:} \xmm image of the cluster highlighting the excess emission smoothed with a $30''$ FWHM Gaussian. The color scale cut off at $1\times10^{-5}$\,cts/s. The emission is located between the two red bars. \textit{Right panel:} \xmm X-ray surface brightness profiles extracted in the direction of the excess emission (orange sector in the left panel) and in neighboring sectors east and west (white sectors in the left panels).}
    \label{fig:sb_enhacement}
\end{figure*}

\subsection{Discontinuity properties}

We define $C$ to be the density ratio across an edge/discontinuity.
\begin{equation}
C = \frac{n_{\rm e, post}}{n_{\rm e,pre}},
\end{equation}
where $n_{\rm e, post}$ and $n_{\rm e, pre}$ are the densities in the region after and before the discontinuity, respectively. 

In case of a shock, assuming Rankine-Hugoniot jump conditions \citep{Landau1959}, we can relate $C$ to the Mach number $\mathcal{M}_{\rm X}$ as:
\begin{equation}
     \mathcal{M}_{\rm X,\rho}= \sqrt{ \frac{2 C}{\gamma + 1 - C (\gamma - 1)}},
\label{eq:M_density}
\end{equation}
where $\gamma = 5/3$ is the ratio of specific heats for monatomic gas.

Using the jump in temperature between the inside and the outside of the identified edges can help discriminate between two possible origins. On the one hand, if the edge is associated with a shock front, the temperature jump is a proxy of the Mach number, as well:
\begin{equation}
    \frac{kT_{\rm post}}{kT_{\rm pre}} = \frac{(\gamma +1)/(\gamma -1)-C^{-1}}{(\gamma +1)/(\gamma -1)-C},
\label{eq:Tshock}
\end{equation}
where $kT_{\rm post}$ and $kT_{\rm pre}$ are the temperatures in the post-shock (downstream) and pre-shock (upstream) regions, respectively. 

The Mach number derived from the temperature jump takes the following form for $\gamma=5/3$:
\begin{equation}
\mathcal{M}_{\rm X,T} = \sqrt{ \frac{ \left(8 \frac{kT_{\rm post}}{kT_{\rm pre}} -7\right) + \sqrt{\left(8 \frac{kT_{\rm post}}{kT_{\rm pre}} -7 \right)^2 + 15} }{5}}.
\label{eq:M_T}
\end{equation}

On the other hand, assuming that the edges are caused by cold fronts in pressure equilibrium (i.e. constant $n_{\rm e} k T$), the expected temperature ratio can be calculated as:
\begin{equation}
    \frac{kT_{\rm post}}{kT_{\rm pre}} = C^{-1}.
\label{eq:Tcf}
\end{equation}

\section{Cluster properties}
\label{sec:clusterproperties}

\subsection{Morphological features}
\label{sec:morphology}

As shown in Figure~\ref{fig:xray_image}, PSZ2\,G181.06+48.47 has an X-ray morphology elongated along the northeast-southwest direction, indicating a cluster merger process is under way along this axis. The cluster displays two X-ray peaks (highlighted with magenta contours in Figure~\ref{fig:multi}, as well as in the GGM-filtered images in Figure~\ref{fig:um_ggm}) separated by $\sim370$\,kpc and connected through a bridge of emission, likely tracing two distinct subclusters observed post-core passage. The concentration ($c$) and centroid-shift ($w$) are among the most robust and widely used parameters to probe the dynamical state of clusters. The concentration parameter is defined as the ratio of the X-ray integrated fluxes in the cluster core and the surrounding outer region of a galaxy cluster. The centroid shift is measured as the standard deviation of the projected separation between the X-ray peak and the centroid (within an aperture radius 500 kpc). It is computed in a series of circular apertures centered on the cluster X-ray peak. The concentration $c=0.087\pm0.001$ and centroid-shift $w=0.053\pm0.001$ parameters place PSZ2\,G181.06+48.47 among the most disturbed clusters: out of 120 clusters studied by \citet{Lovisari2017} only five had both a lower concentration and a higher centroid-shift.

Both subclusters are compact, with the southern subcluster being brighter (see Figure~\ref{fig:xray_image} and~\ref{fig:um_ggm}). The morphology is reminiscent of typical binary, post-core-passage merging clusters, such as CIZA\,J2242.8+5301, PSZ2\,G096.88+24.18, or RXC\,J1314.4-2515 \citep{Golovich2019}. The southern subcluster is about 1.3 times more luminous in the 0.5--2\,keV band than the northern one, supporting a major merger scenario. The spectroscopically-confirmed SDSS DR18 galaxies have an elongated distribution, which matches the merger axis inferred from the X-ray morphology. By cross-matching with the Pan-STARRS imaging and SDSS DR18 spectroscopy (see Figure~\ref{fig:multi}), we identify no magnitude gap ($\Delta m_{\rm R band} = 0$) between the two brightest cluster members. The positions of both BCGs are consistent with the locations of their corresponding X-ray peaks (middle panel in Figure~\ref{fig:multi}).

Between the two subclusters, two symmetrical concave ``bays" border the bridge of emission (see Figure~\ref{fig:um_ggm}). Inside the eastern bay, we detect a loop of emission towards the east of the bridge connecting the subclusters, which is identifiable as an area of very strong gradient (from west to east) in the unsharp-masked and GGM-filtered images. The bays are reminiscent of similar features found in other merging clusters, for example, A\,2256 and A\,746 \citep[e.g.][]{Ge2020, Rajpurohit2024}. Bays are thought to be caused by Kelvin-Helmholtz instabilities driven by the velocity shears between the two merging subclusters \citep[e.g.][]{ZuHone2011b, Walker2017}.

The unsharp-masked and GGM-filtered images reveal that the surface brightness drops more slowly over a distance of $\sim75$\,kpc immediately south of the southern core (see Figure~\ref{fig:um_ggm}). The X-ray emission trailing the southern core towards the south-west is also visible directly in the surface brightness map (see Figure~\ref{fig:sb_enhacement}). We extracted surface brightness profiles in a $30^\circ$-wide sector along the ``trail" of emission, as well as in $30^\circ$-wide sectors to its immediate east and west (see the right panel in Figure~\ref{fig:sb_enhacement}). The enhancement is particularly evident beyond $4.5\arcmin$, where the emission is 1.2--2.9 times above neighboring sectors. The emission could either be associated with a small subcluster, or, given its alignment with the merger axis, more likely stripped gas associated with the merger.

We also detect a clump of emission, $\sim65$\,kpc in extent, toward the east of the northern subcluster. The clump is clearly detected as a relatively compact peak both in the X-ray maps (Figure~\ref{fig:xray_image}) and in the unsharp-masked and GGM-filtered images (Figure~\ref{fig:um_ggm}), and is not associated with any point source or background cluster. While the emission could be associated with a subcluster, given that the clump is offset from any significant dark matter clump \citep[see WL map in][]{PSZ_WL}, its nature is uncertain.

\begin{deluxetable}{c c c c}[!thb]
\tablecaption{Global and subcluster properties.\label{tab:kt500}}
\tablehead{
 & $kT_{500}$ & $L$ & Band \\
          & (keV) & ($10^{44}$\,erg\,s$^{-1}$) & (keV) \\}
\startdata
\multirow{3}{*}{Global} & \multirow{3}{*}{$3.62^{+0.15}_{-0.07}$}  & $1.56^{+0.02}_{-0.01}$ & $0.1-2.4$ \\
   & & $0.95^{+0.02}_{-0.01}$  & $0.5-2.0$ \\
   & & $2.77^{+0.01}_{-0.01}$ & $0.01-100$ \\ \midrule
\multicolumn{4}{c}{Subclusters ($<30''$)}  \\ \midrule
N & $3.73^{+0.25}_{-0.23}$ & $0.19\pm0.01$ & $0.01-100$\\
S & $4.67^{+0.28}_{-0.26}$ & $0.26\pm0.01$ & $0.01-100$\\ \midrule
\multicolumn{4}{c}{Subclusters ($<45''$)} \\ \midrule
N & $3.93^{+0.18}_{-0.18}$ & $0.39\pm0.01$ & $0.01-100$ \\
S & $4.42^{+0.17}_{-0.17}$ & $0.48\pm0.01$ & $0.01-100$  \\ \midrule
 \multicolumn{4}{c}{Subclusters ($<60''$)$^\dagger$} \\ \midrule
N & $4.03^{+0.17}_{-0.15}$ & $0.59\pm0.01$ & $0.01-100$ \\
S & $4.10^{+0.16}_{-0.15}$ & $0.70\pm0.01$ & $0.01-100$ \\
\enddata   
\tablecomments{$^\dagger$At this radius, the areas of the two subclusters overlap.}
\end{deluxetable}

\subsection{Global properties and cluster mass estimation}
\label{sec:global}

The global temperature obtained by jointly fitting \xmm MOS and PN data is $kT_{500}=3.62^{+0.15}_{-0.07}$\,keV (see Table~\ref{tab:kt500}). The global luminosities within $r_{\rm 500,SZ}$ calculated within a range of energy bands are also summarized in Table~\ref{tab:kt500}. PSZ2\,G181.06+48.47 has comparatively a lower temperature and luminosity than samples of clusters studied with \xmm observations \citep[e.g.][]{Pratt2009, Lovisari2020}.

Multiple $M_{500}-kT_{\rm X}$ scaling relations have been presented in the literature, derived from different cluster samples. For example, \citet{Lovisari2020} considered the effect non-cool core clusters have on scaling relations and derived scaling relations using a representative sample of 120 \planck clusters, which included both relaxed and disturbed clusters. The \planck selection and the \xmm-derived X-ray properties should provide a good comparison to our cluster. The presence of two subclusters introduces uncertainty in excising any potential cool core, so we use the relation derived using global, core-included temperatures. Adopting the relation for merging clusters, as appropriate for the highly disturbed PSZ2\,G181.06+48.47 cluster, we obtain a mass of $M_{500,X}=2.32^{+0.29}_{-0.25}\times 10^{14} \; \mathrm{M_{\odot}}$. Using the \citet{Lovisari2020} relation for the entire cluster sample results in a slightly higher, but consistent value ($2.46^{+0.22}_{-0.16}\times10^{14}\;\mathrm{M_{\odot}}$). The resulting $r_{\rm 500,X}$ estimated from the disturbed cluster scaling relation is $0.87\pm0.04\,\mathrm{Mpc}$.

The X-ray derived mass is lower than the \planck mass ($M_{\rm 500,SZ}=(4.2\pm0.5)\times10^{14}\;{\rm M_{\odot}}$), a difference which is significant at the $3.3\sigma$ level. The discrepancy is likely to be even higher, since shock heating can boost the temperature and bias-high temperature-derived masses \citep[as found in A\,2108 by][]{Schellenberger2022}. \citet{Planck2016} cautions that the uncertainties provided represent a lower limit on the real uncertainty, as they do not include all known sources of errors. Another contribution to the discrepancy could be the different centers used in the two measurements (e.g. the midpoint between the subclusters was used for the X-ray measurement, while the \planck measurement is centered closer to the main peak in the south) or the inclusion in the mass estimation of the SZ signal from any of the background clusters identified in the field. As expected, $r_{\rm 500,X}$ is also significantly lower than the \planck value of $r_{\rm 500,SZ}=1.06\pm0.05\,\mathrm{Mpc}$ at the $3.0\sigma$ level.

\citet{PSZ_WL} derive a WL mass for the system of $M_{\rm 500,WL}=2.90^{+0.75}_{-0.69}\times10^{14}\; \mathrm{M_{\odot}}$ by fitting two superpositioned Navarro-Frenk-White halos \citep{NFW1997}. The WL mass is in excellent agreement with the X-ray derived mass. The X-ray and WL-derived masses place the PSZ2\,G181.06+48.47 among the lowest mass clusters to host double radio relics. Out of the 30 known double radio relic systems (see Section~\ref{sec:comparison} for details), only seven have masses $\lesssim3\times10^{14}\,{\rm M_{\odot}}$.

\begin{figure}[!t]
    \centering
    \includegraphics[width=0.48\textwidth]{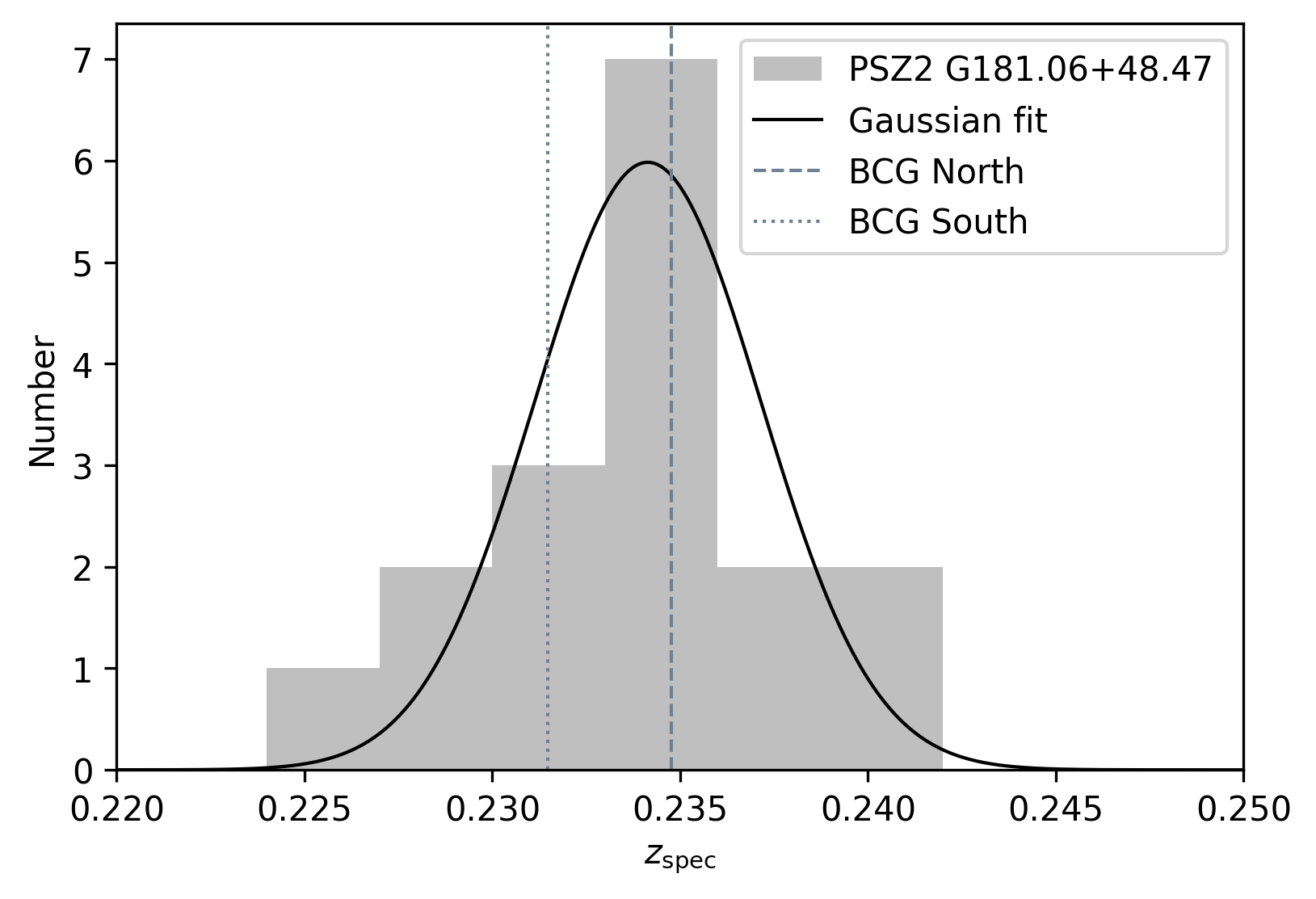}
    \caption{Histogram of SDSS DR18 redshifts within $r_{\rm 500,SZ}$ of the cluster center, with a clear peak at the cluster redshift $z\sim0.234$. The redshifts of the two BCGs are also marked.
}
    \label{fig:zhist}
\end{figure}

SDSS DR18 spectroscopy unveils an overdensity of galaxies at $z\sim0.24$ in the vicinity of the cluster (see Figure~\ref{fig:zhist}). By modeling the limited spectroscopy (17 galaxies) within $r_{\rm 500,SZ}$ of the cluster center with a Gaussian distribution, we estimate the cluster redshift as $z=0.2335\pm0.0035$, with a velocity dispersion of $\sim1290$\,km\,s$^{-1}$ and an equivalent dynamical mass of $2.5\times10^{15}$\,M$_\odot$ \citep[following scaling relations from][]{2013ApJ...772...47S}. \citet{Li2023} have shown that dynamical masses for unrelaxed clusters can be approximately 50\% higher than those derived from X-ray data. Given the disturbed nature of the cluster, this mass estimation is higher than the cluster mass estimated from the gas properties (X-ray temperature and SZ signal). The spectroscopic velocities of the two BCGs ($z_{\rm N}=0.234759 \pm 0.000046$ and $z_{\rm S}=0.231484 \pm 0.000036$) are consistent within the velocity distribution calculated for the cluster. However, the line-of-sight (LOS) velocity difference might indicate that the subclusters are moving towards or away from one another at 982\,km\,s$^{-1}$ along the LOS, consistent with an ongoing merger with a LOS component. The high LOS difference between the two BCGs in PSZ2\,G181.06+48.47 could further inflate the dynamical mass estimate, as was found in other clusters such as Abell\,115 \citep{MKim2019}.

\subsection{Properties of the subclusters}
\label{sec:subclusters}

\begin{figure*}[!thbp]
    \centering
    \includegraphics[width=\textwidth]{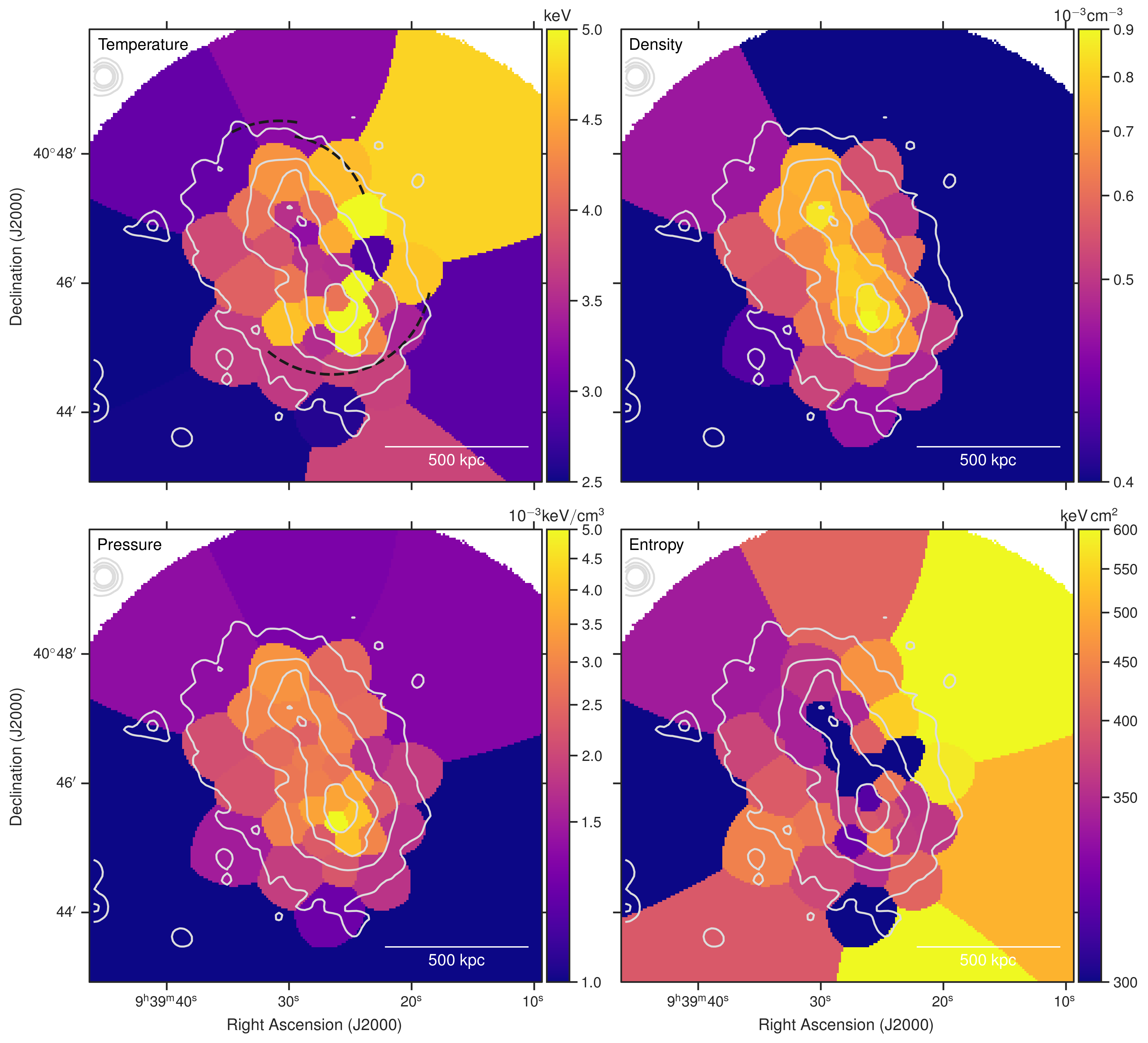}
    \caption{Temperature, density, pressure, and entropy maps focusing on the central $7\arcmin\times7\arcmin$ area of PSZ2\,G181.06+48.47. X-ray surface brightness contours are drawn in gray at $[4, 8, 12, 16]\times10^{-8}\,{\rm cts/s}$ and the edges are highlighted with black dashed curves. The temperature is shown in a linear scale, while the other maps are shown with a square-root stretch.}
    \label{fig:thermo}
\end{figure*}

The mass ratio between two subclusters is a key element for understanding the merger scenario. In the case of PSZ2\,G181.06+48.47, the two cores are located very close in projection to each other and the gas has been heavily mixed in the middle, as indicated by the entropy map (see Section~\ref{sec:thermo} and Figure~\ref{fig:thermo} for further details), which poses challenges to estimating the individual subcluster masses.

We estimate the mass ratio by modeling the subcluster temperatures, as the least biased and lowest scatter method \citep[e.g.][]{Mantz2016, Lovisari2020}. Separate temperature measurements (within $30''$, see Table~\ref{tab:kt500}) of the two subclusters reveal that the brighter, southern subcluster is also hotter than the fainter, northern subcluster by a factor of $kT_{\rm S}{:}kT_{\rm N} = 1.25^{+0.11}_{-0.10}$. When using larger radii of $45''$, the ratio drops to $1.13\pm0.07$, which could indicate a highly concentrated, peaked temperature distribution for the southern subcluster, with a hotter center, and/or a cooler core in the northern subcluster surrounded by possibly higher temperatures (see also Section~\ref{fig:thermo}). We apply the same mass-temperature scaling relation from \citet{Lovisari2020} to obtain mass ratios for the two subclusters of $1.43^{+0.26}_{-0.25}$ and $1.21^{+0.18}_{-0.18}$, for temperatures measured within $30''$ and $45''$ respectively. Considering the possible limitations of the small extraction region in probing the large-scale temperature behavior of each subcluster, we also employ the $M_{2500} \propto T^{1.65\pm0.06}$ relation derived from a sample of 216 cluster, groups, and galaxies by \citet{Babyk2023}. This results in very similar mass ratios of $1.45^{+0.21}_{-0.19}$ and $1.22\pm{0.13}$ for temperatures measured within $30''$ and $45''$ respectively, in line with the scaling relations from \citet{Lovisari2020}. 

Note that while the total mass measured by the WL method \citep{PSZ_WL} is in excellent agreement with the X-ray-derived mass, the mass ratios between the subclusters slightly differ. Both the temperature and luminosity ratios of the two subclusters suggest that the southern subcluster is about 1.2--1.4 times more massive than the northern subcluster, while the WL finds a mass ratio that is closer to 1:3 ($3.0\pm1.3$). On the one side, the proximity between the two subclumps might result in some projection effects that slightly affect the temperature estimate. On the other side, the WL mass map reveals extensions towards the southeast and northwest of the southern subcluster core, but no corresponding feature is found in the X-ray map, possibly indicating subgroups that were stripped of gas during their infall into the PSZ2\,G181.06+48.47 system \citep{Eckert2014, Ebeling2017}.

\begin{deluxetable*}{c c c c c c }[!thbp]
\tablecaption{Surface brightness profiles across the discontinuities fit with a single and broken power law model and temperatures in the pre- and post-discontinuity regions.}
\label{tab:disc}
\tablehead{ & N edge & NW  edge & S edge & NE relic & SW relic}
\startdata
 & \multicolumn{5}{c}{Broken power-law model} \\ \cline{2-6}
$r_{\rm f}$ & $1.61\arcmin\pm0.07\arcmin$ & $1.39\arcmin\pm0.10\arcmin$ & $ 1.53\arcmin\pm0.07\arcmin$ &  $6.53\arcmin$$\ddagger$ & $5.30\arcmin$$\ddagger$ \\
$C$  & $1.46\pm0.15$ & $1.52\pm0.18$ & $1.45\pm0.08$ & $1.00\pm0.13$$^\diamond$ & $1.00\pm0.17$$^\diamond$ \\
$\chi^2_{\rm red}$$^\dagger$ & 0.98 & 1.65 & 1.21 & 0.80 & 0.80 \\ \hline
 & \multicolumn{5}{c}{Single power-law model} \\ \cline{2-6}
$\chi^2_{\rm red}$ & 2.78 & 3.13 & 22.91 & 0.80 & 0.85 \\ \hline
 & \multicolumn{5}{c}{Temperatures} \\ \cline{2-6}
$kT_{\rm post}$ & $5.43^{+1.27}_{-0.92}$ & $4.23^{+0.38}_{-0.35}$ & $4.02^{+0.16}_{-0.15}$ & --$^*$ & --$^*$ \\
$kT_{\rm pre}$ & $2.77^{+0.54}_{-0.39}$ & $4.01^{+0.45}_{-0.46}$ &
$3.86^{+0.27}_{-0.29}$ & --$^*$ & --$^*$ \\
\enddata
\tablecomments{$^\dagger$Reduced $\chi^2$, obtained by dividing the $\chi^2$ by number of degrees of freedom. $\ddagger$Fixed. $^\diamond$Constrained to values above 1. $^*$Reliable temperature measurements were not possible in regions beyond $r_{500,P}$, where the S/N is low.}
\end{deluxetable*}

\subsection{Thermodynamic properties}
\label{sec:thermo}

In line with its classification as a major merger, the cluster displays rich, complex thermodynamic structures. The temperature, density, pressure and entropy maps can be found in Figure~\ref{fig:thermo}.

In the upper left panel of Figure~\ref{fig:thermo}, we show the temperature map with the identified edges in the N, NW and S highlighted. Figure~\ref{fig:kT_uncertainty} contains the relative uncertainties associated with the temperature map. The temperatures in the central region of the cluster are constrained to within 10--16\%. The northern core and the region directly to its south along the bridge are the coolest areas in central region of the cluster. The temperatures along the bridge seem to increase from the northern to the southern core, from a low of $\sim3.2$\,keV to a peak of 5.5\,keV in the region south of the south core. Toward the south, the southern core is enveloped in an arc of higher temperature gas. Compared to the southern core itself, this nearby area is hotter, and has lower density and higher entropy, possibly indicating evidence for shock heating.

The highest densities coincide with the location of the subclusters, as implied by the X-ray map and the positions of the BCGs. The density map (upper right panel in Figure~\ref{fig:thermo}) reveals a high density region in the cores of the two subclusters and the connecting bridge, flanked by areas of lower density. Outside of the cores, the density drops more sharply along the merger axis (NE of the NE core and SW of the SW core), than in the perpendicular direction. Pressure peaks correspond to both subcluster cores, with the southern core possibly showing peak pressures $\sim60-70$\% above the northern core (see the lower left panel in Figure~\ref{fig:thermo}). 

The low-entropy southern core is embedded in a comparatively higher entropy region than the northern core (lower right panel in Figure~\ref{fig:thermo}). In addition, a bridge of low entropy gas connects the two subclusters, with the lowest entropies found in the northern core. Note however that the relative BCG velocities suggest a LOS component to the merger, which could imply that the connecting bridge also has a LOS component. If all of that gas has the same temperature, the measured temperature would not be affected, but the normalization and inferred density will be driven up. The combined effect would artificially raise the projected pseudo-entropy. Simulations have shown that, in the absence of non-gravitational effects, the low-entropy, dense, cold gas settles to the center of relaxed galaxy clusters \citep[e.g.][]{Voit2002, McCarthy2008, Power2014}. The low-entropy gas corresponding to each of the cores indicates that the two subclusters might have been cool-core, relaxed clusters before the merger event. While the merger likely did not fully disrupt either of the two subcluster cores, there is stronger evidence for disruption in the southern core. Towards the west, perpendicular to the merger axis, regions of possibly hot ($\sim5.1$\,keV) and cold ($\sim2.9$\,keV) plasma are located next to each other. The cold area has higher densities, lower pressures, and significantly lower entropies. Such low entropy plumes can be formed most cleanly shortly ($\sim500$\,Myr) after head-on collisions between clusters of similar masses \citep{ZuHone2011b}. The features observed here could be related to the southeast-northwest extension observed in the WL mass map \citep{PSZ_WL}, possibly indicating a third-body collision experienced by the southern subcluster region, or sloshing activity potentially triggered by the core passage.

\begin{figure*}[!tbh]
    \centering
    \includegraphics[width=0.49\textwidth]{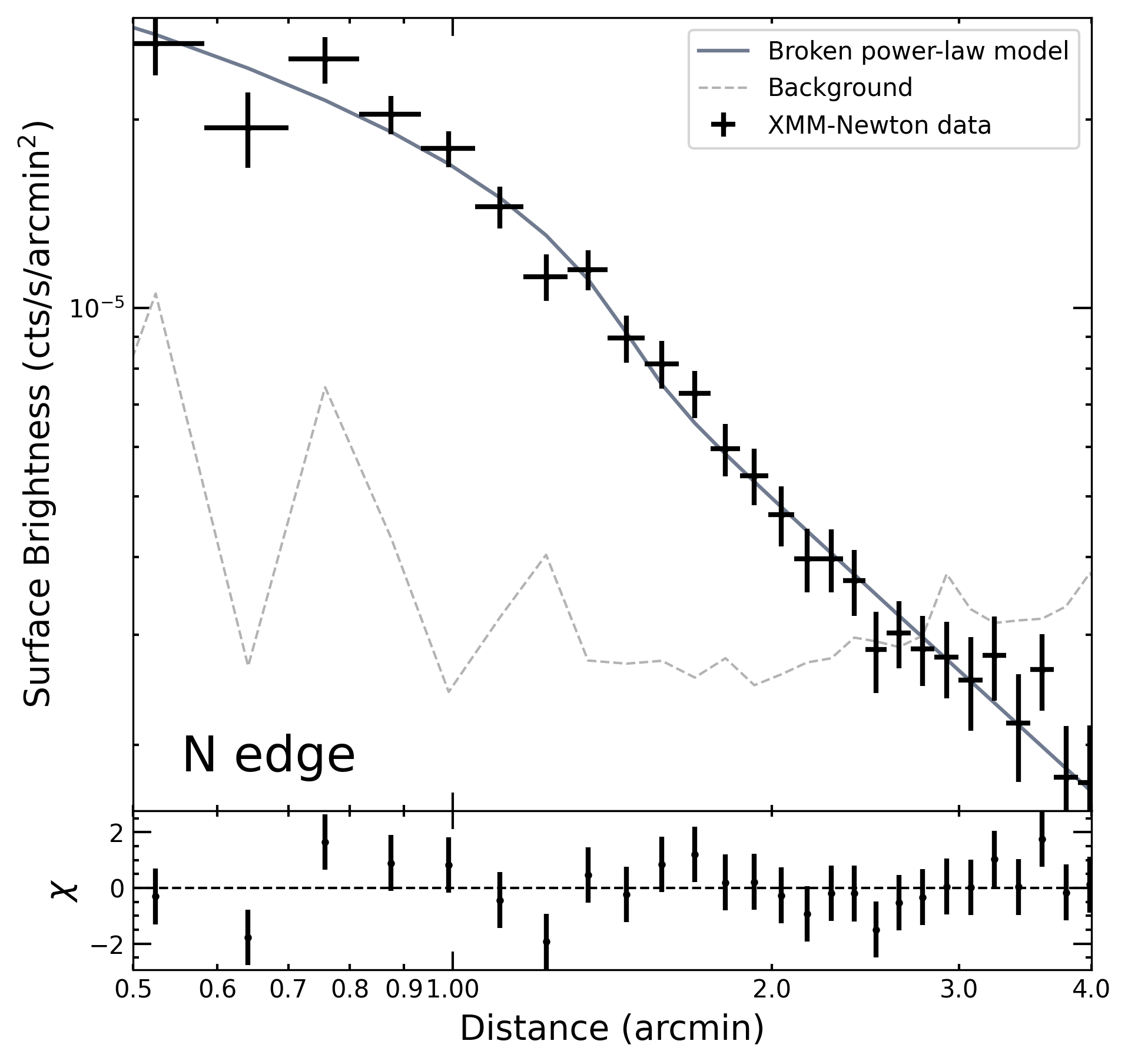}
    \includegraphics[width=0.49\textwidth]{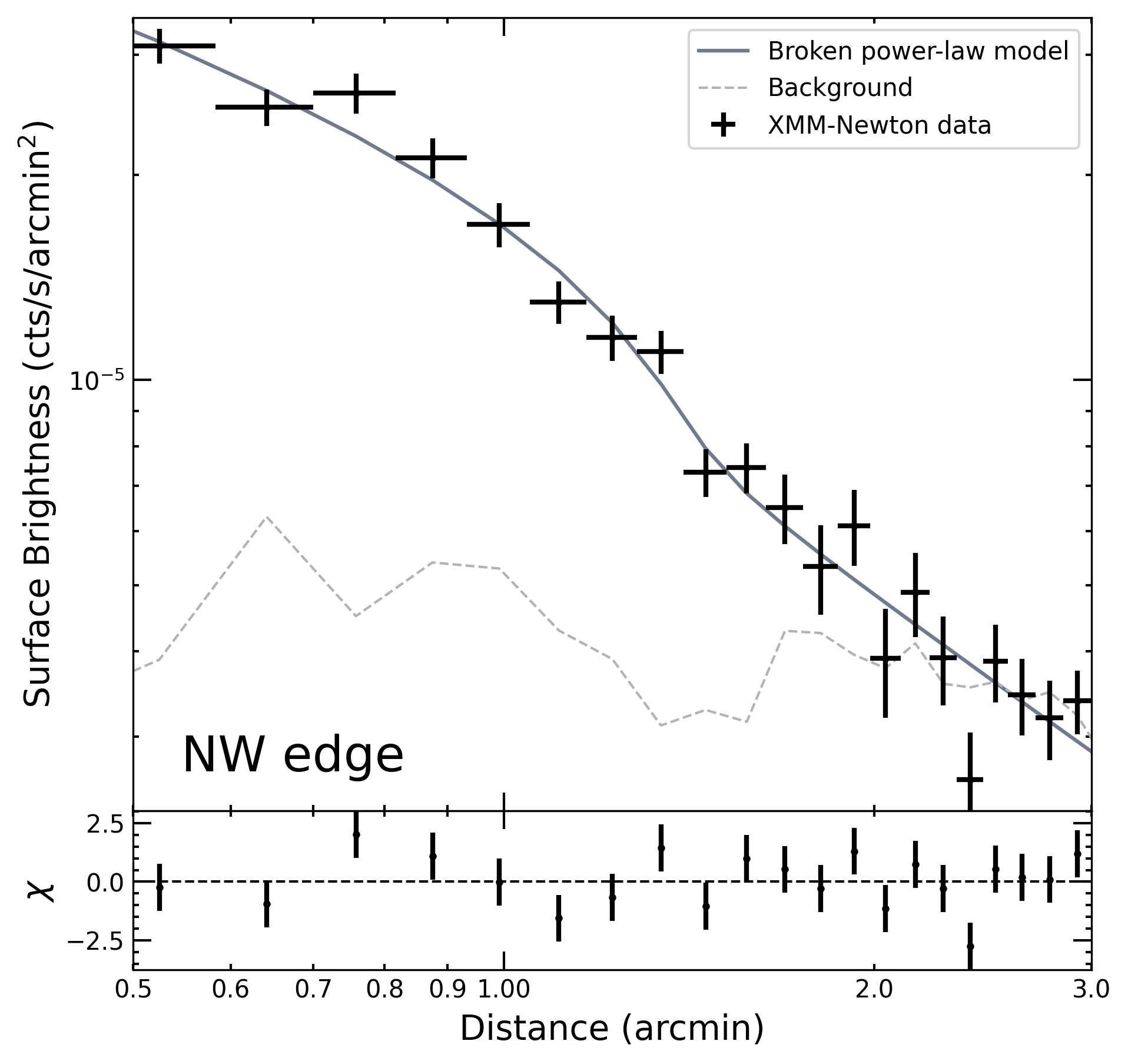}
    \includegraphics[width=0.49\textwidth]{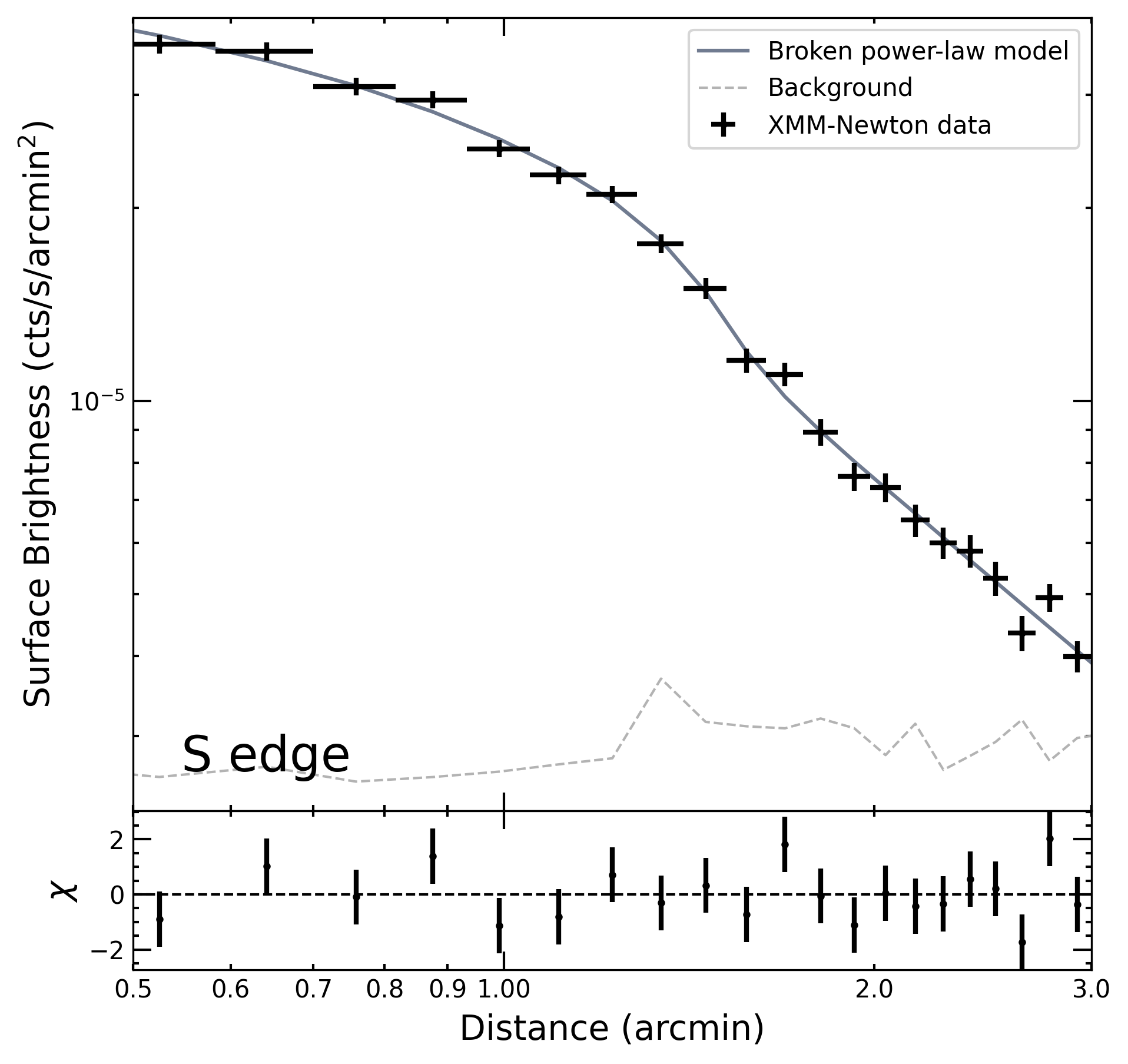}
    \caption{\xmm 0.5--2\,keV X-ray surface brightness profiles near the N (upper left panel), NW (upper right panel) and S edge (lower panel) with the corresponding best-fit broken power-law models. The dashed lines mark the non-X-ray background that has been subtracted. The bottom panels show the fit residuals. The corresponding profiles and the location of the edges are marked in Figure~\ref{fig:extract_regions}.}
    \label{fig:sb_disc}
\end{figure*}

\section{Search for discontinuities}
\label{sec:discontinuities}

\subsection{Discontinuities close to the cluster core}
\label{sec:discontinuities:central}

We take a two-pronged approach to identify X-ray discontinuities. Firstly, through a visual search of the unsharp mask and GGM filtered images, we identify two candidate edges close to the cluster core, one in the northwest (NW) and another in the south (S), highlighted with dashed curves in Figure~\ref{fig:um_ggm}. Secondly, we confirm the presence of these edge candidates and search for other edges (if present) in radial surface brightness profiles within the central ($<4\arcmin$) areas of the cluster. 

The \xmm surface brightness analysis confirms the NW and S edges as significant discontinuities and unveils another potential discontinuity (labeled N) towards the north of the cluster (see Figure~\ref{fig:sb_disc}). The sectors used to extract the surface brightness profiles are shown in Figure~\ref{fig:extract_regions}. The two sectors for the NW and the S edges follow the curvature of the visually identified edges as marked in Figure~\ref{fig:extract_regions}. Note that varying the properties of the extraction sectors (such as the opening angle, fitting radius range) does not significantly alter the fit results. Surface brightness profiles extracted along the discontinuities are shown with the best-fit models in Figure~\ref{fig:sb_disc}, with fit summaries listed in Table~\ref{tab:disc}. The upstream and downstream temperatures are also listed in Table~\ref{tab:disc}.

All three discontinuities in the central region are aligned roughly with the N-NE--S-SW merger axis inferred from the X-ray morphology of the system and are located around $1.5\arcmin$ from their sector centers. This places the discontinuities at $1\arcmin-1.3\arcmin$ or a projected physical distance of $\sim225-290$\,kpc from the center of the nearest subcluster and $\sim270-500$\,kpc from the system center. The N and NW discontinuities are detected at a S/N of $\sim3$. The NW edge has the highest compression, a compression factor $C=1.51\pm0.18$. However, the S edge is detected at the highest confidence level of $5.6\sigma$ ($C=1.45\pm0.08$). In all three cases, the broken power-law model provides a good fit to the data with $\chi^2_{\rm red}$ close to 1, while the single power-law model provides a comparatively poorer fit. This is especially true for the high S/N S edge, where the single power-law model $\chi^2_{\rm red}\sim23$, while the broken power-law model provides an excellent fit with $\chi^2_{\rm red}\sim1.2$.

The temperature ratios for the NW and S edge are consistent with unity, within the uncertainties, precluding us from securely classifying them as cold fronts or shocks. The temperature ratio of $1.96^{+0.60}_{-0.43}$ found across the N edge prefers a shock classification at the $>2\sigma$ level.

We also explored the \chandra surface brightness profiles in the sectors defined above. The \chandra observations confirm the presence, location and jump strength of the discontinuities in the central area as measured from the \xmm observations. While the \chandra measurements are in agreement with \xmm in terms of location and jump strength of the discontinuities, we note the agreement is within the large error bars on the \chandra fit parameters, which are caused by the lower S/N of the \chandra observations.

Under the assumption that the density edges are due to shocks, we can estimate the X-ray Mach number $\mathcal{M}_{\rm X,\rho}$ using equation~\ref{eq:M_density}. We can also derive the Mach number $\mathcal{M}_{\rm X,T}$ from the temperature jump from equation~\ref{eq:M_T}. All three edges would correspond to mild shocks (see Table~\ref{tab:mach_number}), with $\mathcal{M}_{\rm X,\rho}<1.4$ and broadly consistent $\mathcal{M}_{\rm X,T}$. Interestingly, the temperature-derived Mach number for the S edge is lower than the density-based Mach number, with a significance of $2.5\sigma$. By contrast, $\mathcal{M}_{\rm X,T}=1.91^{+0.48}_{-0.35}$ suggests the N edge could be a slightly stronger shock than implied by $\mathcal{M}_{\rm X,\rho}=1.31\pm0.11$ ($1.6\sigma$ difference). No corresponding radio features are detected at the location of these edges. However, low acceleration efficiency is expected for low Mach number shocks when particles are injected from the thermal pool \citep[$\lesssim10^{-4}$ for $\mathcal{M}<3$, e.g.][]{Kang2007, Kang2013}. Regardless, the lack of radio emission does lend more support to a scenario where, at least the NW and S edges, are possibly cold fronts, rather than shocks.

The inner discontinuities in PSZ2\,G181.06+48.47 are reminiscent of those found in clusters such as CIG\,0217+70 \citep{Tumer2023}, CIZA\,J2242.8+5301 \citep{Ogrean2014} and A\,746 \citep{Rajpurohit2024}. While the discontinuities in A\,746 and CIZA\,J2242.8+5301 are located at beyond $>0.5 r_{500}$ \citep{Rajpurohit2024, Ogrean2014}, in CIG\,0217+70, \citet{Tumer2023} find three shock fronts within $0.5 r_{500}$, which they attribute to a secondary stage of merging.

\begin{deluxetable}{c c c}[!t]
\tablecaption{X-ray Mach number $\mathcal{M}_{\rm X}$ for features in the PSZ2\,G181.06+48.47 system, calculated from the density and temperatures jumps. \label{tab:mach_number}}
\tablehead{Feature  & $\mathcal{M}_{\rm X,\rho}$ & $\mathcal{M}_{\rm X,T}$}
\startdata
    N edge  & $1.31\pm0.11$ & $1.91^{+0.48}_{-0.35}$ \\
    NW edge & $1.36\pm0.13$ & $1.06\pm0.16$ \\
    S edge & $1.31\pm0.06$ & $1.04\pm0.09$ \\
    NE shock & $<1.43$$^\dagger$ & -- \\
    SW shock & $<1.57$$^\dagger$ & -- \\
\enddata
\tablecomments{$5\sigma$ upper limit.}
\end{deluxetable}

\begin{figure*}[!tbh]
    \centering
    \includegraphics[width=0.49\textwidth]{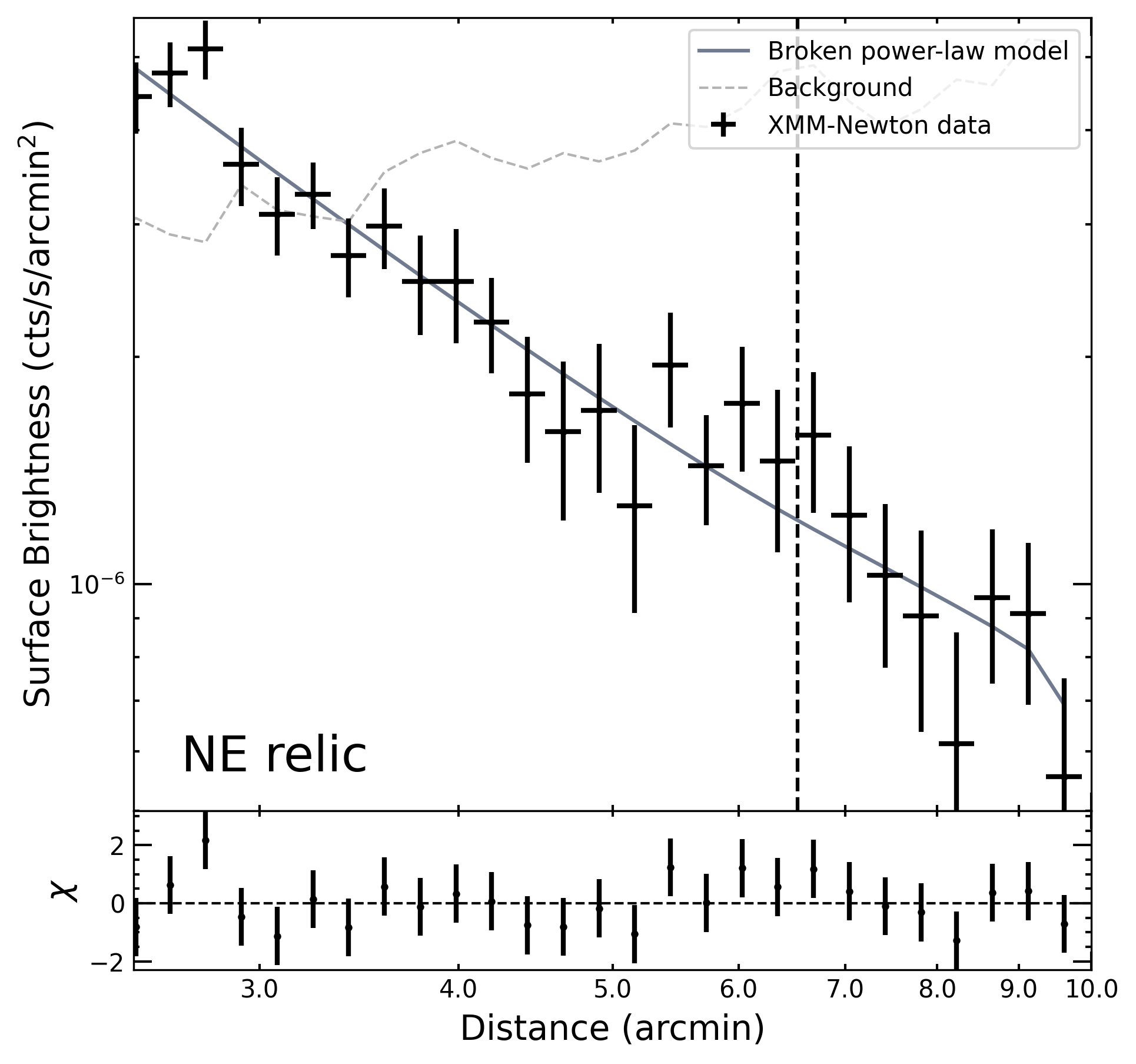}
    \includegraphics[width=0.49\textwidth]{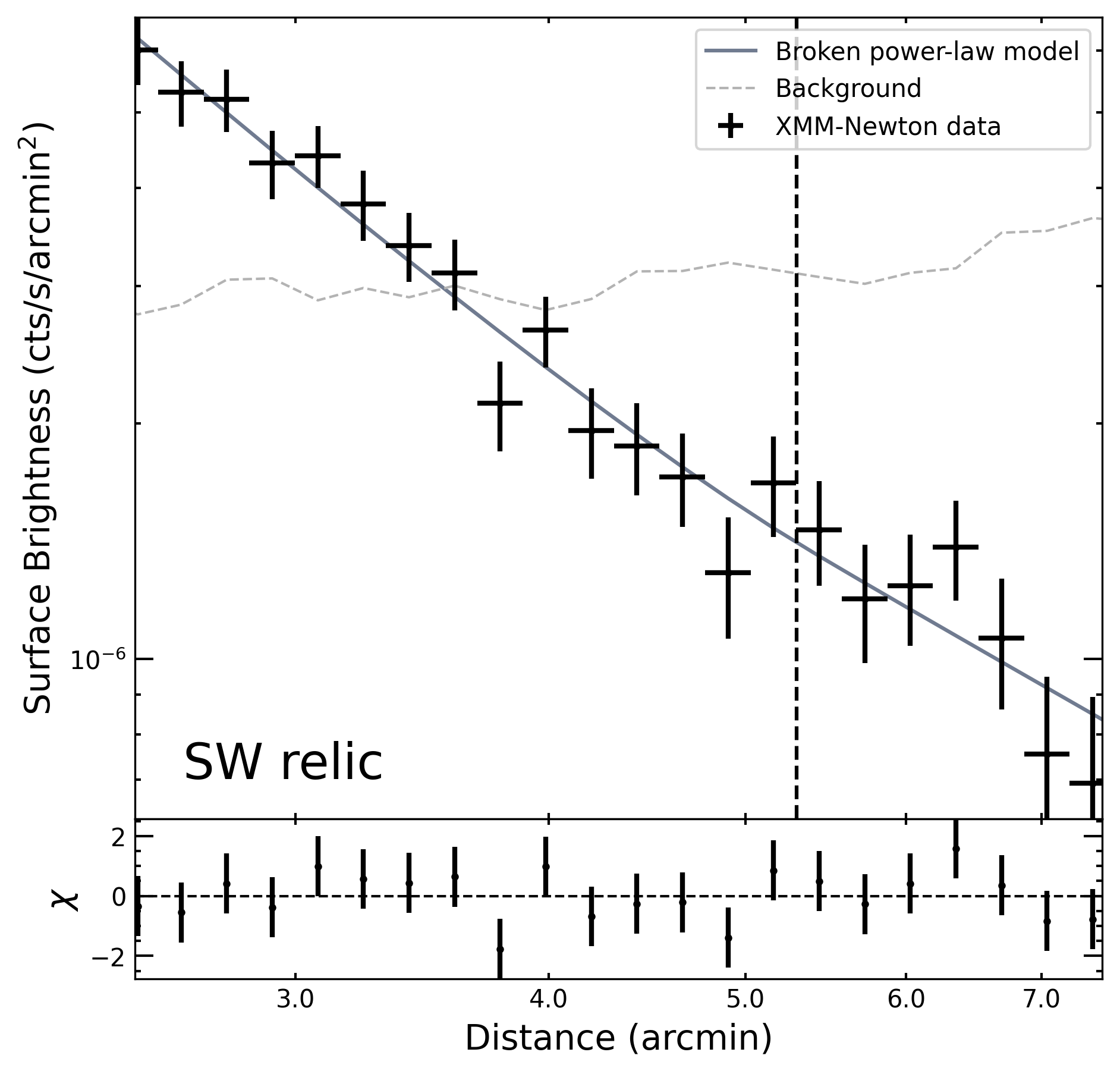}
    \caption{
    \xmm 0.5--2.0\,keV X-ray surface brightness profiles across the NE (left panel) and SW radio (right panel) relics, extracted from the corresponding sectors shown in Figure~\ref{fig:extract_regions}. No significant shock was discovered at the relic locations, indicating the shocks have X-ray derived Mach numbers of at most $\sim1.5$.}
    \label{fig:sb_relics}
\end{figure*}

\subsection{Discontinuities at the radio relic positions}
\label{sec:discontinuities:relics}

We investigated the \xmm profiles near the diffuse radio emission in the NE and SW, to search for potential shock fronts associated with the radio relics. The extraction regions are marked in Figure~\ref{fig:extract_regions}. The shapes of the two sectors were especially designed to follow the curvature of the relics, in order to search for shock candidates. 

As shown in Figure~\ref{fig:sb_relics}, while no clear density jumps were detected in these regions, we can use a broken-power law model to place limits on the X-ray Mach number of the shocks. The position of the shocks was fixed to the outer edge of the radio relic, where the putative location of the shock is expected to fall (see Figure~\ref{fig:extract_regions}). For the NE relic, from the fit compression ratio of $1.00\pm0.13$, we place a $5\sigma$ upper limit on the Mach number of $\mathcal{M}_{\rm X,\rho}<1.43$, while for the SW relic the compression ratio $1.00\pm0.17$ yields a $5\sigma$ limit of $\mathcal{M}_{\rm X,\rho}<1.57$. Note that if there are significant projection effects at play caused by the merger being offset from the plane-of-the-sky, the intrinsic density-derived Mach number might be higher \citep[e.g.][]{2023A&A...679A.161O}. We further consider the potential uncertainty in the X-ray derived Mach numbers due to the assumed ratio of specific heats $\gamma = 5/3$. For fully relativistic gas, the specific heat ratio can approach $4/3$, resulting in $\mathcal{M}_{\rm X}<1.38$ and $<1.50$ for the NE and SW shocks, respectively, which suggests the shocks may be even weaker. Since the extraction regions for the relics are located partially beyond $r_{\rm 500,P}$, the low S/N regime impacts our ability to make temperature measurements. We are, therefore, unable to constrain the temperatures in the pre- and post-shock regions at the relic locations. 

As described in \citet{PSZ_radio}, a significant spectral gradient has been observed for the SW relic in the spectral index map, which rules out its classification as a double-lobed radio galaxy, and indicates a radio relic nature. The integrated spectral index implies a high Mach number shock $\mathcal{M}_{\rm R}=4.8^{+2.3}_{-0.9}$ under the assumption of diffusive shock acceleration (DSA), significantly above the upper limit for the X-ray Mach number $\mathcal{M}_{\rm X, \rho}<1.57$. 

Radio Mach numbers typically lie above the corresponding X-ray Mach numbers (e.g., \citealt{vanWeeren2016a, Hoang2018, Stuardi2019} and \citealt{vanWeeren2019} for a review). From the X-ray perspective, the estimated Mach number could be lowered by a number of observational factors, for example, the mixing of plasma with different properties within the large sectors required by the low number counts at the cluster periphery and projection effects \citep[e.g.][]{2023A&A...679A.161O}. Further, the merger axis of the cluster is likely tilted at an angle $\geq45\degree$ with respect to the plane-of-the-sky \citep[e.g. velocity difference between BCGs and radio considerations from][]{PSZ_radio}, which could result in unaccounted-for projection effects, such as mixing of plasma along the LOS, as well as a ``blurring" of the X-ray discontinuity. As further discussed in \citet{PSZ_radio}, the discrepancy is most likely explained by the intrinsic differences in which part of the Mach number distribution the X-ray and radio trace, with X-ray observations probing the average of the Mach number distribution, while the radio probes the high Mach number tail \citep[e.g.][]{Wittor2021, Paola2021}.

\section{Comparison with other double radio relic systems}
\label{sec:comparison}

There are well-documented correlations among relic properties, such as radio power, largest linear scale (LLS), and correlation between relic properties and cluster-related properties, such as cluster mass and relic distance to the cluster center \citep{deGasperin2014, Jones2023, Duchesne2024}. Probing these relations is difficult because relics are intrinsically rare (10\% occurrence rate) and their typical location at the cluster periphery, where X-ray counts are low, makes them challenging to observe. A recent study on PSZ2 clusters shows that more relics await discovery in clusters with masses $M_{500}\lesssim5\times10^{14}$\,M$_\odot$ \citep{Jones2023}.

\begin{figure*}[!hbt]
    \hspace*{-5.87cm}\includegraphics[width=0.318\textwidth]{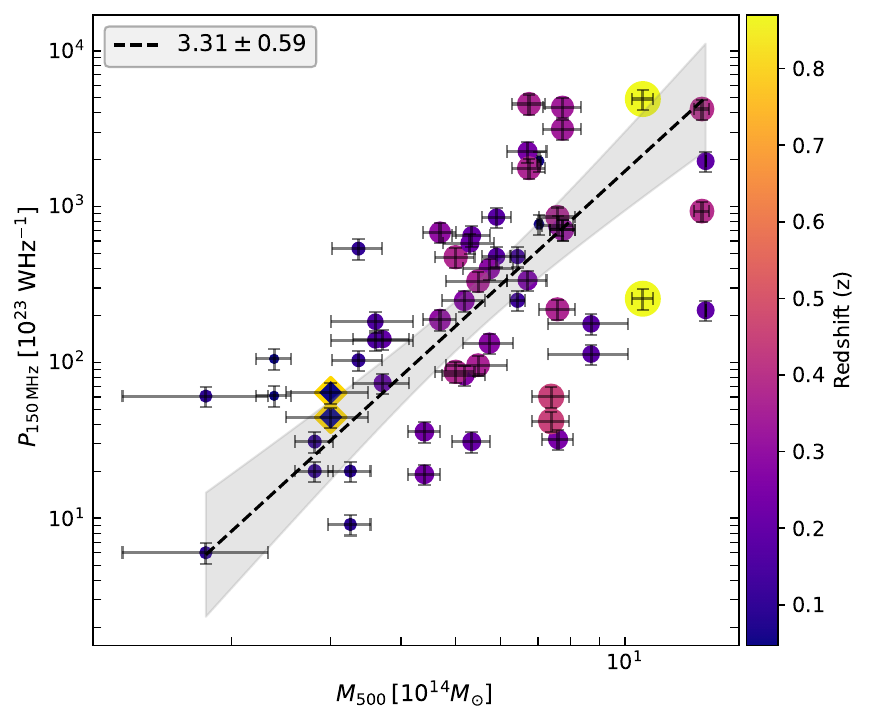} \\
    \centering
    \includegraphics[width=1.0\textwidth]{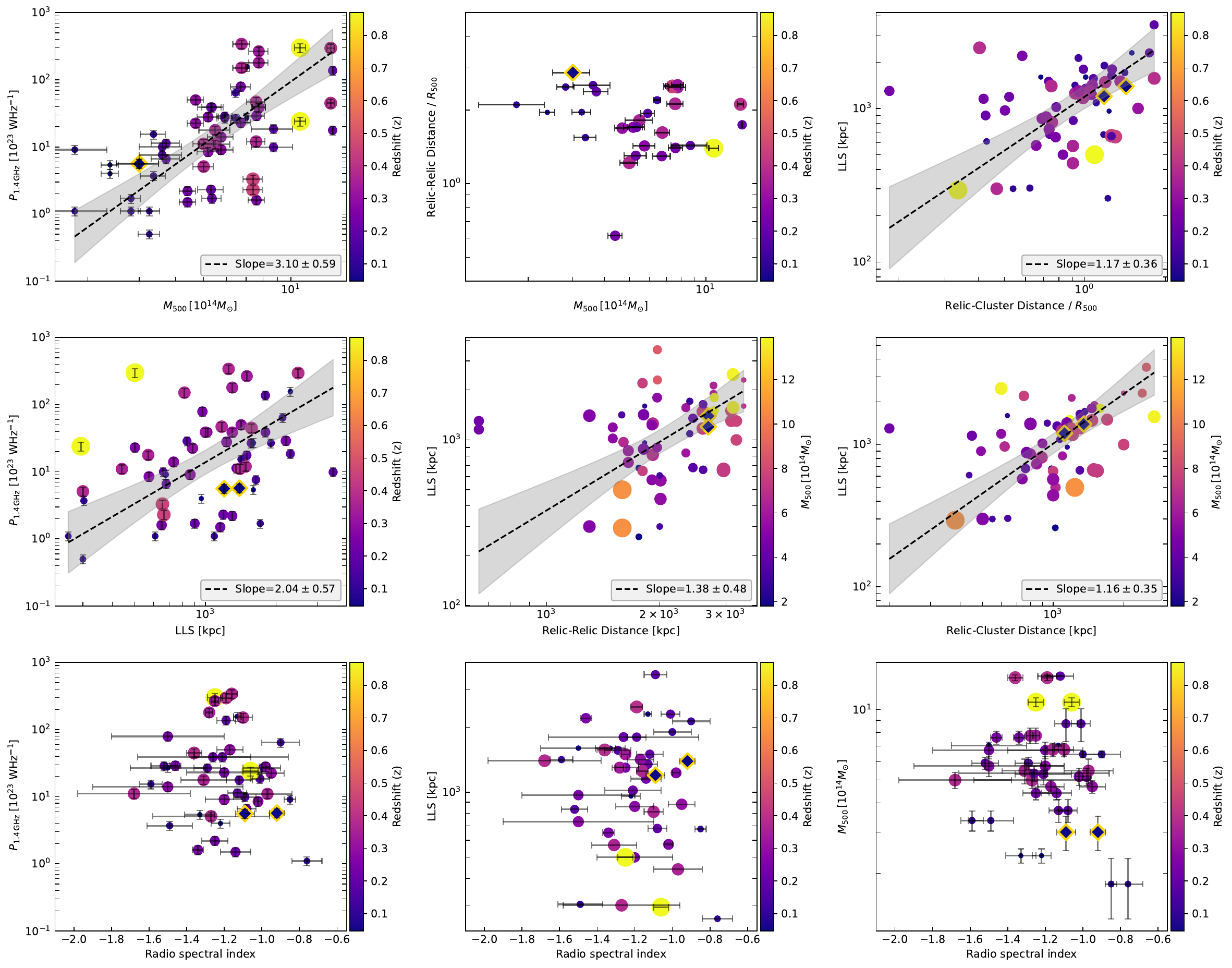}
    \caption{Known relations for double radio relics and comparison to the relics in PSZ2\,G181.06+48.47. To the best of our knowledge, all the double relics to date are included, along with their updated radio properties (see Table~\ref{table:double_radio_relic}). The black dashed lines indicate the orthogonal BCES regression fit, with the 95\% confidence interval shown as a gray shaded area where the correlation is significant ($p<0.05$). The diamond-shaped data points (with yellow border) represent the relics in PSZ2\,G181.06+48.47. Relics without spectral index information were excluded from the bottom row panels.}
    \label{fig:relic_properties}
\end{figure*}

\cite{deGasperin2014} observed a positive correlation between the 1.4\,GHz radio power of double relics, the cluster mass, and relic LLS. To perform a comprehensive analysis of the scaling relations, we systematically searched the literature for all double radio relics identified to date. Double relics are thought to trace mergers in the plane-of-the-sky, thus making them less affected by projection effects and ideal for studying scaling relations \citep[e.g.][]{Golovich2019, vanWeeren2019}. In Table~\ref{table:double_radio_relic}, we compile data for 30 galaxy clusters hosting double relics, including a total of 60 relics. This sample includes 12 new systems containing double radio relics compared to \citet{Jones2023} and \citet{deGasperin2014}. The clusters span a mass range of  $1.7 \times 10^{14} M_{\odot} < M_{500} < 14 \times 10^{14} M_{\odot}$  and a redshift range of  $0.04 < z < 0.87$. We estimated radio power as: 
\begin{equation}
P_{1.4 \, \text{GHz}} = 4\pi D_L^2 S_{\nu{0}} (1 + z)^{-(1 + \alpha)},
\end{equation}
where $D_L$ represents the luminosity distance to the source,  $\alpha$ denotes the spectral index applied for the k-correction, and  $S_{\nu_0}$  refers to the flux density measured at $\nu_0 = 1.4\,\rm GHz$. The error in the radio power was estimated by assuming a 15\% uncertainty in the radio flux densities for the relics. 

The 1.4\,GHz radio power for each relic was calculated using updated spectral index values. Where three or more flux density measurements were available, we determined the spectral index by fitting the data with a power-law. Where only two measurements were available, we assumed the spectral index implied by a power-law passing through both points. For the small subset (14/60) of relics where only a single flux density was known, we assumed a spectral index $\alpha = -1.3$, which is typical for relics. If the 1.4\,GHz measurement was available, it was used as the basis of the power estimate. Where no measurement was available, we extrapolated the 1.4\,GHz flux density using the spectral index. The resulting plot is shown in Figure~\ref{fig:relic_properties}. 

As reported by \citet{PSZ_radio}, the 1.4\,GHz radio power of the NE and SW relics is $(5.7 \pm 0.8) \times 10^{23}\, \rm W\,Hz^{-1}$ and $(5.6 \pm 0.5) \times 10^{23}\, \rm W\, Hz^{-1}$, respectively. The PSZ2\,G181.06+48.47 relics occupy the lower end of the $P_{1.4}$ vs. $M_{500}$ relation and fit well with other known double relics (see Figure~\ref{fig:relic_properties}). We find that PSZ2\,G181.06+48.47 has the largest separation between the relics scaled with $r_{\rm 500,SZ}$, despite its low mass. The relics in PSZ2\,G233.68+36.14  have the second largest separation.

\subsection{Double relic scaling relations}

We calculated Spearman's rank correlation coefficient to determine the significance of each relation. The null hypothesis, which assumes no correlation between the variables X and Y, was considered rejected when $p<0.05$. If the null hypothesis was rejected, we fit the data with BCES orthogonal regression\footnote{\url{https://github.com/rsnemmen/BCES}} and calculated the 95\% confidence interval for the best-fit. We opted for the orthogonal fit to ensure consistency with other relic studies \citep{deGasperin2014, Jones2023}. We report the p-values and best-fit parameters in Table~\ref{tab:beces_fitting}. 

\begin{deluxetable}{lcc}[!t]
\tablecaption{Spearman rank correlation coefficient p-values and best-fit parameters for correlations in Figure\,\ref{fig:relic_properties}. \label{tab:beces_fitting}}   
\tablehead{Correlation  & p-value & Slope}
\startdata
$\rm P_{1.4\,GHz}-M_{500}$ &$6\times10^{-8}$&$3.10\pm0.59$\\
$\rm P_{1.4\,GHz}-LLS$ & $0.002$ &$2.04\pm0.57$ \\
$\rm P_{1.4\,GHz}-\alpha$ & 0.47 &  $-$  \\
$\rm D_{RR}/R_{500}-M_{500}$ & 0.55 &  $-$   \\
$\rm LLS-D_{RR}$& 0.002 &  $1.38\pm0.48$ \\
$\rm LLS-\alpha$& 0.73 &   $-$   \\
$\rm LLS-D_{RC}/R_{500}$& 0.005 &  $1.17\pm0.36$ \\
$\rm LLS-D_{RC}$& 0.0001 &  $1.16\pm0.35$ \\
$\rm M_{500}-\alpha$ & 0.36& $-$   \\
\enddata
\tablecomments{We fit the data, using BCES orthogonal regression method, only when the null hypothesis was rejected.}
\end{deluxetable}

Figure~\ref{fig:relic_properties} reveals a strong correlation ($p = 6 \times 10^{-8}$) between double relic 1.4\,GHz power and cluster mass, providing compelling evidence to reject the null hypothesis. A discernible trend suggests that high-redshift clusters tend to be more massive, while relics with lower radio power are predominantly found in nearby clusters. The median redshift of our sample is 0.233, so about half the sample is located above this redshift. Since seven of the 10 clusters at $z>0.3$ were discovered by \planck and the other three by X-ray observations, the mass limit evolves with redshift, naturally resulting in a bias toward massive clusters at higher redshifts \citep[e.g.][]{Planck2016}. Despite the slightly different cosmology and the expanded sample which doubles the size of our sample compared to \citet{deGasperin2014} and improves upon it with updated values for radio power, spectral index, and related properties, the slope of the radio power at 1.4\,GHz vs. cluster mass relation is consistent within the errors: our value of $3.10 \pm 0.59$ compared to their value of $2.83 \pm 0.39$. When the 14 relics without spectral index measurements are excluded, we obtain a slope of $2.78\pm 0.58$. The radio power of relics is theoretically expected to scale with the cluster mass following the relation $P\propto M^{5/3}$ \citep{deGasperin2014}. However, both our and the \citet{deGasperin2014} observations indicate a different mass dependency, with $P_{1.4\, \mathrm{GHz}} \propto M^{3.10 \pm 0.59}$, which is steeper than expected theoretically (i.e, $M^{5/3}$), at the $2.4\sigma$ level. \citet{deGasperin2014} suggest that this discrepancy arises from parameters that depend on cluster mass which, in turn, drive particle acceleration efficiencies or magnetic field strengths. It is worth noting that \citet{Jones2023}, using their sample and that of \citet{deGasperin2014}, reported a much steeper dependency of $P_{150\,\mathrm{MHz}} \propto M^{5.2 \pm 1.2}$ indicating a significant difference between the relations at 150\,MHz and 1.4\,GHz. To investigate this, we also compare the power-mass scaling relations of double relics in our sample with those of \cite{Jones2023} at 150\,MHz. To do this, we converted the 1.4\,GHz radio power to 150\,MHz using the spectral index. For the 14 relics without spectral index information, we assumed a constant value of $\alpha = -1.3$ (top, left panel in Figure~\ref{fig:relic_properties}). In contrast to \cite{Jones2023}, we find that the mass dependence on both 150\,MHz and 1.4\,GHz radio power is strikingly similar. Specifically, the measured correlation slope is $3.31 \pm 0.59$. We emphasize that, when excluding relics lacking spectral index information, the measured slope remains consistent, with a value of $3.19 \pm 0.61$. 

We also find a significant correlation between radio power at 1.4\,GHz and LLS ($p=0.002$). A similar trend was previously hinted at by studies from \citet{Bonafede2012} and \citet{Jones2023}. If we exclude double relics without spectral index information, the null hypothesis is rejected at $p=0.007$. This implies that relics with higher radio power are generally associated with larger linear sizes, suggesting that larger relics are likely more powerful. However, no significant dependence on redshift is observed in this case.  
 
Recently \citet{Jones2023} reported a positive correlation between the LLS of relics and their projected cluster distance ($D_{\rm RC}$) as a fraction of the cluster $r_{500}$. In line with this study, we also find that the larger relics tend to be located farther from the cluster center ($p=0.005$), with a best-fit slope of $1.17\pm0.36$. The farther out, the steeper the gas density profile and the more important the ``runaway'' shock effect would be, possibly explaining the observed correlation.

We also examined the relationship between the relic-relic distance ($D_{\rm RR}$), expressed as a fraction of the cluster $r_{500}$, and cluster mass. The null hypothesis was not rejected ($p = 0.55$), indicating no correlation. However, it is noteworthy that the relics in PSZ2\,G181.06+48.47 exhibit the largest separation, especially when considering the system's low-mass, making the system unique. 

We find significant correlations between LLS vs. $D_{\rm RR}$ as well as $D_{\rm RC}$ (see Figure~\ref{fig:relic_properties}). The shock surfaces could thus increase in size with greater distance from the cluster center, resulting in larger relics farther from the cluster center, consistent with prior findings by \citet{Bonafede2012}, \citet{deGasperin2014} and \citet{Jones2023}. Our best fit slopes are $\mathrm{LLS}\propto  D_{\rm RC}^{1.16\pm0.35}$ and $\mathrm{LLS}\propto D_{\rm RR}^{1.38\pm0.48}$. The latter slope is more or less consistent with \citet{deGasperin2014}, namely $\mathrm{LLS} \propto D_{\rm RR}^{2.13 \pm 0.25}$. Conversely, the slope for LLS vs. $D_{\rm RC}$ is consistent with the results of \cite{Jones2023} at 150\,MHz, supporting their conclusion that LLS does not vary significantly at lower frequencies.

We find that the null hypothesis is not rejected for the relationships between radio power, LLS, and cluster mass versus radio spectral index, indicating no significant correlation. This is consistent with the previous studies of the radio power-spectral index relation \citep{Bonafede2012, deGasperin2014}. The other two relations have not been explored in the literature, precluding any direct comparison. The absence of a correlation with the spectral index is expected because relics typically exhibit spectral indices within the narrow range of $-0.9$ to $-1.3$, whereas cluster masses, redshifts, and relic LLS span much broader ranges. 

\section{The Merger Scenario}
\label{sec:mergerscenario}

PSZ2\,G181.06+48.47 is a low-mass post-core passage system, resulting from the merger of two subclusters along the N-NE--S-SW axis. The post-merger scenario is strongly supported by the bridge of cool, low-entropy gas connecting the two subcluster cores which indicates gas stripping during the first passage. Although the dynamical mass is biased high, the X-ray, WL, and \planck mass estimates ($(2.5-4.2)\times 10^{14} \; \mathrm{M_{\odot}}$) confirm that PSZ2\,G181.06+48.47 is one of the lowest-mass systems to host a double radio relic system.

While double relic systems are typically assumed to be less affected by projection effects, PSZ2\,G181.06+48.47 might not conform to this expectation. While the X-ray data do not directly constrain the orientation of the merger in the plane-of-the-sky, the symmetric nature, comparable sizes, and thin width of the two radio relics and the similar properties of the subclusters would, at first glance, suggest that the merger is occurring in the plane-of-the-sky. However, based on a radio polarization and color-color analysis, \citet{PSZ_radio} ruled out the possibility that both relics are seen close to edge-on. Instead, \citet{PSZ_radio} presented compelling evidence that the system could be tilted at an angle as high as $\sim45^\circ$. The northern BCG is also located at a higher LOS velocity ($\Delta v\sim980$\,km\,s$^{-1}$) compared to the southern one, lending further support that the merger axis is tilted with respect to the plane-of-the-sky. \citet{Takizawa2010} provided insights from hydrodynamical simulations indicating that the dynamical/virial mass can be more than twice the actual mass when observing along the collision axis at $t=1.33$\,Gyr after the first core passage. The possibility of a significant impact parameter, especially along the LOS, is further discussed later on.

The $\sim370$\,kpc projected separation between the X-ray peaks of the two subclusters is dwarfed by comparison to the $\sim2.6$\,Mpc projected separation between the two radio relics, complicating the interpretation of how double relics can form at such large separation, where the density of particles available for acceleration is very low \citep[e.g.][]{Brunetti2014, Botteon2020}. As discussed above, if the merger axis is not in the plane-of-the-sky, the two relics would be physically located even farther apart from each other \citep[$\sim3.7$\,Mpc][]{PSZ_radio}. The limited time available for evolution would then present new challenges in explaining how relics propagate to such large distances. \citet{PSZ_radio} propose that the relics are associated with shock re-acceleration of particles previously accelerated through, for example, active galactic nuclei, or re-acceleration through multiple sequential shocks. The relics follow typical scaling relations, having typical LLSs and powers, with their large cluster-centric being the main remarkable feature. 

Given the low mass of the cluster and the large separation of the relics, the relics could be ``runaway shock" candidates. \citet{CZhang2019} discuss two phases of evolution for merger shocks: an initial phase, where the shock is ``driven" by the infalling subcluster, and the ``detached" phase, where the direction and speed of the shock can deviate significantly from the behavior of the infalling subcluster. The resulting ``runaway" shocks are expected to be more often found at large cluster-centric radii $>r_{500}$, consistent with the combination of a low-mass cluster and large separations observed in PSZ2\,G181.06+48.47 \citep{CZhang2019}. 

We consider four possible scenarios below: 1) an head-on outgoing merger, 2) a post-apocenter infalling merger, 3) a post second pericenter passage, outgoing merger and 4) an offset merger.

The alignment of the relics with the X-ray axis would be a natural outcome of head-on outgoing merger. The small projected separation between the two subclusters ($\sim370$\,kpc) would restrict the potential timescales for the merger to a very young, post-core passage merger observed soon after core passage. If we are witnessing the outgoing phase, after the first core passage, the time left for the detached shock waves to reach the cluster outskirts (their projected radii) is \citep{Sarazin1986}:
\begin{equation}
\frac{v}{\rm km\,s^{-1}} < \mathcal{M} c_{s} = 1480 \mathcal{M} \sqrt{\frac{T}{10^{8}\,\rm K}}.
\end{equation}
Using the measured upper limits of the Mach number from the density jumps, the minimum required travel times are 0.78\,Gyr and 1.07\,Gyr for the SW and NE shocks. The constraint is slightly relaxed to 0.45\,Gyr and 0.55\,Gyr if the temperature-derived Mach numbers are used. An outgoing scenario cannot simultaneously explain the small core separation and the large relic-relic distance. The tail of emission behind the southern core as well as the inner discontinuities would also be challenging to explain in such a scenario. It is, therefore, unlikely that PSZ2\,G181.06+48.47 is undergoing an outgoing phase.

The tension between the small X-ray core distance and the large relic separation would be eased if the cluster is an older merger, seen post-apocenter, with the subclusters in the approaching phase. The excess X-ray emission towards the S-SW of the system could be associated with material stripped from the southern subcluster. This material could be trailing the southern subcluster if the system is observed post-apocenter, after the first core-passage, as the two subclusters are falling back into each other. A post-core passage merger with a significant LOS component would also explain the discrepancy between the dynamical mass and X-ray mass. If the system is in a late-stage, post-apocenter merger, the ``runaway'' phase of the shocks would also occur. As the subclusters are moving back towards each other, the shocks would travel further out, with increasing Mach number, as the subclusters fall back in, thus explaining the large separation and brightness of the radio arcs. The radio emission in ``runaway" shocks is expected to trace a narrow shell, in line with the narrow radio width of the two relics in PSZ2\,G181.06+48.47. A similar scenario was proposed, for example, for the Coma cluster, in which \citet{Lyskova2019} suggest that the NGC\,4839 group originated from the northeast, crossed the cluster core, has already reached apocenter and, after passing the apocenter, reversed its direction to re-enter the core. However, the shock generated during first passage continued moving towards the cluster outskirts, creating the radio relic. The set of symmetric shocks at inner and larger radii could be explained via ``N-waves'' \citep{CZhang2021a}, which are expected to form closer to the cluster center, in the rarefied regions left behind the ``runaway'' shock. The ``runaway" shocks are expected to be associated with strong radio emission, while the strong inner shocks are more easily detectable in X-ray observations \citep{CZhang2021a}. The set of observational parameters for our inner shocks and the pair of relics match well this scenario. While N-waves are most clearly identified in minor mergers, with mass ratios of 1:10, simulations show can also be observed in major mergers such as PSZ2\,G181.06+48.47 \citep{CZhang2021a}.

If the subclusters are seen in the outgoing phase, after the second core passage, the small core distance would be easily understood. A post-apocenter scenario would naturally lead to X-ray edges close to the core, which could be formed more recently during the second encounter of the two cores, similar to the scenario proposed for CIG\,0217+70 by \citet{Tumer2023}. The inner shocks would then be driven by the second infall of the subclusters post-apocenter. However, compared to the post-apocenter scenario, the time added to cross the cores would likely lead to even larger separations between the relics, as discussed through simulations by \citet{PSZ_WL}. Furthermore, the bridge would likely be completely destroyed and mixed in as the two subclusters pass through each other for a second time as indicated in simulations by \citet{ZuHone2011b}.

A fourth possible scenario could be that the clusters are not in a head-on collision. For example, a small, non-zero impact parameter merger could most readily produce a longer-lived bridge of low-entropy emission that connects the two subclusters \citep{ZuHone2011b}. Simulations by \citet{Sheardown2019} have shown that seemingly head-on collisions can be explained by a ``sling-shot'' scenario. For simulated cluster mergers with a mass ratio of $\sim$1:1, the arc-shaped sling-shot tails appear symmetric on both sides of the cluster in the early stage, as observed in galaxy groups NGC\,7618 and UGC\,12491 \citep{Roediger2012}. Shortly after the apocenter passage, both subclusters slow down, turn around and begin to infall again. The bow shock detached from the slowing-down subcluster will continue moving away from the cluster center, creating the relics in the outskirts. One piece of evidence in support of this scenario is the tail of X-ray emission at the south of the southern subcluster. While the S/N in that region is too low to make any conclusive morphological analysis, the trail seems to align with the merger axis, with no evidence for the bowed distribution expected for a sling-shot tail candidate, implying that any bending would have to be along the LOS, consistent with the evidence of an inclined merger axis and an off-axis collision where the impact parameter vector is along the LOS direction.

Given the observational evidence, we thus propose that the cluster is most likely in a late-stage merger, seen after the first apocenter passage, as the subclusters are falling back towards each other. 

\section{Conclusion}
\label{sec:conclusion}

We explored new \chandra and \xmm X-ray observations of the low-mass merging cluster PSZ2\,G181.06+48.47, which hosts two spectacular symmetrically located double relics. We reveal that the cluster is very cool ($kT_{500}=3.62^{+0.15}_{-0.07}$\,keV) and has a mass, $M_{500,X}=2.32^{+0.29}_{-0.25}\times10^{14} \;{\rm M_{\odot}}$, significantly lower than that previously measured by \planck, but consistent with the estimate from the WL \citep{PSZ_WL}. Among the lowest mass clusters ever discovered to host double radio relics, the system is a major merger along the N-NE--S-SW direction, between two clearly-identified subclusters with a mass ratio 1.2--1.4.

The cluster stands as one of the most disturbed clusters in the \planck sample and displays a complex morphological and thermodynamic structure, which includes concave bays perpendicular to the merger axis, enhanced emission trailing the southern subcluster and evidence for shock heating. The two subclusters are connected through a low-entropy bridge, indicating a partial disruption of the cool-cores and stripping of gas during the merger.

We discover a set of three, weak ($\mathcal{M}_{X,\rho}\sim1.3-1.4$) shocks located within $<500$\,kpc of the system center and aligned with the merger axis. The lack of significant density jumps provides an upper limit of $\mathcal{M}_{\rm X,\rho}\lesssim1.6$ for the NE and SW relic shocks, significantly lower Mach numbers compared to estimations from the radio data \citep{PSZ_radio}, suggesting an inclination of the system with respect to the plane-of-the-sky. 

Through a careful analysis of the existing literature, we compiled a sample of all known double radio relic systems and revised scaling relations by adding 12 new systems not included in previous work. We find positive correlations between relics power vs mass, relics power vs LLS, LLS vs relic-cluster distance, and LLS vs relic-relic distance. The radio relic power at 1.4\,GHz scales as $P_{1.4\,\mathrm{GHz}} \propto M_{500}^{3.10 \pm 0.59}$, consistent with previous findings. We find that the relics in PSZ2\,G181.06+48.47 have typical luminosities and sizes, but they are located exceptionally far away from the cluster center, around $r_{200,WL}$, considering the low mass of the cluster. The nature of the diffuse radio sources in PSZ2\,G181.06+48.47 is investigated in \citet{PSZ_radio}.

The existing observational evidence points to a late-stage, post-merger scenario with a significant LOS component, where the two subclusters are seen post-apocenter as they are falling back into each other. The outer shocks produced during the first encounter expand outwards, reaching large radii over time, which would account for the large distance between the two relics. The merger scenario is further explored through simulations in \citet{PSZ_WL}. 

Identifying faint radio relics such as those produced in low-mass clusters, as well as relics with large relic separation has been challenging. However, with the advent of new-generation radio telescopes and surveys, we may be uncovering the ``tip of the iceberg". Perhaps PSZ2\,G181.06+48.47 is representative of a broader range of low-mass systems with diffuse radio emission which will be discovered in the future with more sensitive instruments, all posing a firm challenge to our expectations that particle acceleration would be inefficient at low Mach number shocks. Confirming shocks in these low-mass clusters has been complicated by the associated low S/N regimes at large cluster centric radii. Next-generation X-ray missions such as \textit{Athena} \citep{Nandra2013} and the Hot Universe Baryon Surveyor \citep[HUBS;][]{Cui2020} will transform our understanding of particle acceleration in the low-mass regime.

\section*{Acknowledgments}

We thank the anonymous referee for their useful feedback. We thank Felipe Andrade-Santos, Tim Shimwell and Grant Tremblay for useful discussions. A. Stroe, Z. Zhu, and K. Rajpurohit acknowledge support from the NASA 80NSSC21K0822 and \chandra GO0-21122X grants. A. Stroe gratefully acknowledges previous support from a Clay Fellowship. Z. Zhu and A. Simionescu are supported by the Netherlands Organisation for Scientific Research (NWO). L. Lovisari acknowledges support from INAF grant 1.05.12.04.01. W. Forman acknowledges support from the  Smithsonian Institution, the Chandra High Resolution Camera Project through NASA contract NAS8-03060, and NASA Grants 80NSSC19K0116 and GO1-22132X. M. James Jee acknowledges support for the current research from the National Research Foundation (NRF) of Korea under the programs 2022R1A2C1003130 and RS-2023-00219959. This work is based on observations obtained with \xmm, an ESA science mission with instruments and contributions directly funded by ESA Member States and the US (NASA). This research has made use of observations made by the \chandra X-ray Observatory, and software provided by the \chandra X-ray Center (CXC). LOFAR \citep{Haarlem2013} is the Low Frequency Array designed and constructed by ASTRON.

\vspace{5mm}
\facilities{XMM, CXO, LOFAR, SDSS, Pan-STARRS}

\software{
pyproffit \citep{Eckert2020},
Astropy \citep{astropy2013, astropy2018},
APLpy \citep{aplpy},
Matplotlib \citep{matplotlib},
TOPCAT \citep{topcat},
DS9 \citep{ds9},
CARTA \citep{CARTA2021},
CIAO \citep{CIAO},
SAS}

\bibliography{PSZ_Xray}{}
\bibliographystyle{aasjournal}

\appendix
\restartappendixnumbering

\section{Temperature uncertainty map}

\begin{figure*}[!thbp]
    \centering
    \includegraphics[width=0.5\linewidth]{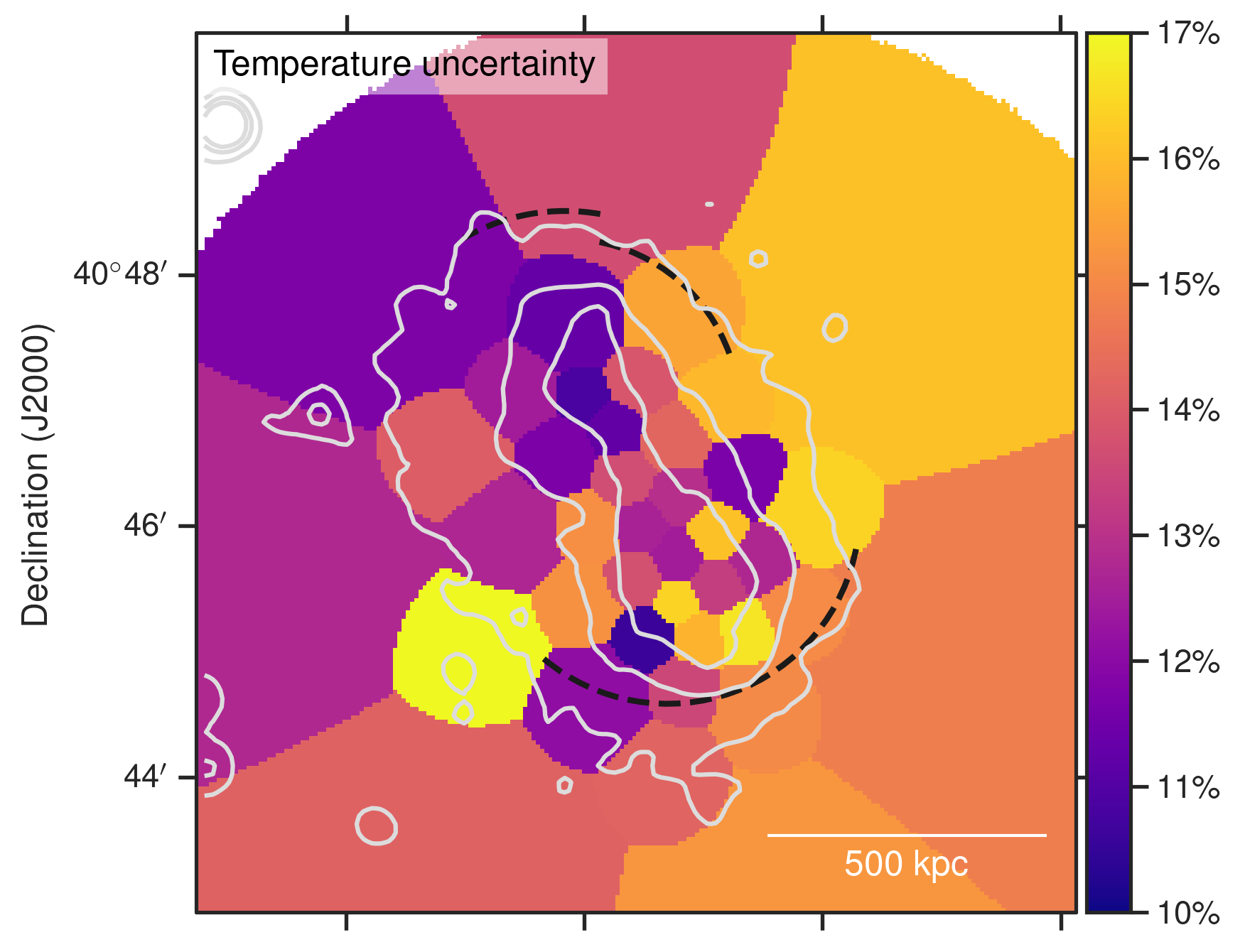}
    \caption{Fractional temperature uncertainty, corresponding to the temperature map shown in Figure~\ref{fig:thermo}.}
    \label{fig:kT_uncertainty}
\end{figure*}

\section{Galaxy clusters hosting double radio relics}

\setlength{\tabcolsep}{3pt}
\setlength{\LTpre}{0pt} % Space before the first table on the first page
\setlength{\LTpost}{0pt} % Space after the last table on the last page
\begin{longtable}{p{4.3cm}*{9}{c}}
\caption{Properties of the 30 known double radio relics.} \label{table:double_radio_relic} \\
\hline \hline
Cluster$_{relic}$  &  $z$  & LLS & $d_{\rm RC}$ & $d_{\rm RR}$& $P_{\rm 1.4GHz}$ & $M_{500}$ &$\alpha$& References \\
 & & (kpc) & (kpc) & (kpc) & ($10^{23}\,{\rm W}\,\rm{Hz}^{-1}$) & ($10^{14}{\rm M}_{\odot}$)&  \\
\hline
\endfirsthead

\multicolumn{9}{c}%
{{\bfseries \tablename\ \thetable{} -- continued from previous page}} \\
\hline \hline
Cluster$_{relic}$  &  $z$  & LLS & $d_{\rm RC}$ & $d_{\rm RR}$& $P_{\rm 1.4GHz}$ & $M_{500}$  &$\alpha$& References \\
 & &(kpc) & (kpc)& (kpc) & ($10^{23}\,{\rm W}\,\rm{Hz}^{-1}$) & ($10^{14}{\rm M}_{\odot}$) & & \\
\hline
\endhead

\hline \multicolumn{8}{r}{{Continued on next page}} \\ \hline
\endfoot

\hline \hline
\endlastfoot
A\,3365$_{E}$ & \multirow{2}{*}{0.093} & 676 & 984 &\multirow{2}{*}{1760} & 9.1 & \multirow{2}{*}{1.80$\pm0.52$$\dagger$} &$-0.85\pm0.03$ &  \multirow{2}{*}{Du21, Kn22} \\
A\,3365$_{W}$ &  & 262 & 1021 & & 1.2 & & $-0.76\pm0.08$ &  \\\hline
A\,2345$_{E}$ & \multirow{2}{*}{0.177} & 1565 & 1304 &\multirow{2}{*}{2348} & 27 & \multirow{2}{*}{5.91$^{+0.37}_{-0.35}$} &$-1.29\pm0.07$& \multirow{2}{*}{Bo09, Ge17} \\
A\,2345$_{W}$ &   & 834  & 1096 & & 29 &  & $-1.52\pm0.07$& \\\hline
A\,1240$_{N}$ & \multirow{2}{*}{0.1948} & 680 & 762 &\multirow{2}{*}{2452} & 6.6 & \multirow{2}{*}{3.70$^{+0.41}_{-0.42}$} & $-1.08\pm0.05$&\multirow{2}{*}{Bo09, Ho18} \\
A\,1240$_{S}$ &  & 1350 & 1039 &&  11.3 &  & $-1.13\pm0.05$&\\\hline
A\,3376$_{E}$ & \multirow{2}{*}{0.0468} & 1600 & 638 &\multirow{2}{*}{1824} &5.4 &  \multirow{2}{*}{2.38$^{+0.16}_{-0.17}$} & $-1.33\pm0.08$&\multirow{2}{*}{Ch23} \\
A\,3376$_{W}$ &  & 960 & 1149 & & 4.0 & &$-1.22\pm0.05$ &\\\hline
A\,3667$_{SE}$ & \multirow{2}{*}{0.0556} & 1600 & 1344 & \multirow{2}{*}{3347}& 27 & \multirow{2}{*}{7.03$\pm0.06$} & $-1.50\pm0.2$&\multirow{2}{*}{dG22} \\
A\,3667$_{NW}$ &  & 2300 & 2000 && 157 & & $-1.13\pm0.02$&\\\hline
RXC\,J1314.4-2515$_{E}$ & \multirow{2}{*}{0.2439} & 500 & 1023 &\multirow{2}{*}{1602}  & 23 & \multirow{2}{*}{6.71$^{+0.53}_{-0.54}$} & $-1.2\pm0.2$&\multirow{2}{*}{St19} \\
RXC\,J1314.4-2515$_{W}$ &  & 970 & 621 &  & 79 &  &$-1.5\pm0.3$ &\\\hline
A\,2146$_{NW}$ & \multirow{2}{*}{0.2339} &1156 &450 &\multirow{2}{*}{660}& 1.5&\multirow{2}{*}{4.04$^{+0.27}_{-0.29}$} & $-1.14\pm0.08$&\multirow{2}{*}{Ho19} \\
A\,2146$_{SE}$ & & 1300& 200& & 2.2& &$-1.25\pm0.07$& \\\hline
A\,521$_{NW}$ & \multirow{2}{*}{0.2475} &650 &1000 &\multirow{2}{*}{1942}&1.61& \multirow{2}{*}{7.26$^{+0.48}_{-0.49}$}&$-1.34\pm0.03$ &\multirow{2}{*}{Sa24}\\
A\,521$_{SE}$ & &2200&750&&29& & $-1.46\pm0.03$&\\\hline
A\,3186$_{NW}$& \multirow{2}{*}{0.127} &1900 & 1500 &\multirow{2}{*}{2782}&26.8&\multirow{2}{*}{6.44$\pm0.20$}& $-1.0\pm0.1$&\multirow{2}{*}{Du21, Kn22} \\
A\,3186$_{SE}$ & &2130  & 1200& &64 & &$-0.9\pm0.1$ &\\\hline
MACS\,J1752.0+4440$_{NE}$ & \multirow{2}{*}{0.366} & 1254 & 1091 & \multirow{2}{*}{1941}&341 & \multirow{2}{*}{6.75$^{+0.44}_{-0.46}$} &$-1.16\pm0.03$ &\multirow{2}{*}{vW12} \\
MACS\,1752.0+4440$_{SW}$ & & 813 & 892 & &151 & &$-1.10\pm0.05$ &\\\hline
CIZA\,J2242.8+5301$_{N}$ & \multirow{2}{*}{0.189} & 1800 & 1585 & \multirow{2}{*}{2793}&137 &   \multirow{2}{*}{$\sim13.9$$\ddagger$} & $-1.19\pm0.05$&\multirow{2}{*}{vW10; St13; DG21} \\
CIZA\,J2242.8+5301$_{S}$ &&1500&1175&&17.7&&$-1.12\pm0.07$ &\\\hline
ZwCl\,2341.1+0000$_{N}$ &\multirow{2}{*}{0.27} & 574 & 738 & \multirow{2}{*}{1907} &8.5&  \multirow{2}{*}{5.18$^{+0.43}_{-0.45}$} & $-1.02\pm0.02$&\multirow{2}{*}{vW09, St22} \\
ZwCl\,2341.1+0000$_{S}$ & & 1230 & 1189 &&27.9 &&$-0.98\pm0.02$ &\\\hline
ZwCl\,1447.2+2619$_{N}$ &\multirow{2}{*}{0.376} & 300& 500 & \multirow{2}{*}{1300}&5.1 &  \multirow{2}{*}{$\sim5$$^{\dagger\dagger}$} & $-1.27\pm0.31$&\multirow{2}{*}{Le22} \\
ZwCl\,1447.2+2619$_{S}$ & & 1400 & 800 &&11.1 &&$-1.68\pm0.30$ &\\\hline
ZwCl\,0008.8+5215$_{E}$ &\multirow{2}{*}{0.104}&1416&944&\multirow{2}{*}{1585}&15.4&\multirow{2}{*}{3.36$^{+0.32}_{-0.34}$}&$-1.59\pm0.06$&\multirow{2}{*}{vW11a}\\
ZwCl\,0008.8+5215$_{W}$ &&303&640&&3.7&&$-1.49\pm0.12$&\\\hline
CIG\,0217+70$_{E}$ & \multirow{2}{*}{0.18} & 3500 & 2500 & \multirow{2}{*}{1972}&9.9 & \multirow{2}{*}{8.7$\pm1.4$$^*$} & $-1.09\pm0.06$&\multirow{2}{*}{Zh20; Ho21}\\
CIG\,0217+70$_W$ & & 2300 & 2400 & &18.5 & &$-1.01\pm0.05$ &\\\hline
ACT-CL\,J0102-4915$_{NW}$ & \multirow{2}{*}{0.87}&500&1237&\multirow{2}{*}{1590}&300&\multirow{2}{*}{10.75$^{+0.48}_{-0.47}$}&$-1.25\pm0.04$ &\multirow{2}{*}{Li14, St22}\\
ACT-CL\,J0102-4915$_{SE}$ & &294&382&&24&&$-1.06\pm0.04$& \\\hline
SPT-CL\,J2032-5627$_{NW}$ & \multirow{2}{*}{0.284}&860&800&\multirow{2}{*}{1650}&9.1& \multirow{2}{*}{5.74$^{+0.56}_{-0.59}$} &$-1.2\pm0.1$ &\multirow{2}{*}{Du21}\\
SPT-CL\,J2032-5627$_{SE}$ & &731&850&& 14&&$-1.5\pm0.1$& \\\hline
PSZ2\,G096.88+24.18$_{N}$ &\multirow{2}{*}{0.30}&880&770&\multirow{2}{*}{1812}&22.5&\multirow{2}{*}{4.69$^{+0.30}_{-0.31}$}& $-0.95\pm0.07$&\multirow{2}{*}{Jo21}\\
PSZ2\,G096.88+24.18$_{S}$ &&1419&1145&&50&&$-1.17\pm0.07$ &\\\hline
PSZ2\,G108.17-11.56$_{NE}$ &\multirow{2}{*}{0.337}&1500&1700&\multirow{2}{*}{3100}&266&\multirow{2}{*}{7.74$^{+0.57}_{-0.60}$}&$-1.25\pm0.02$&\multirow{2}{*}{dG15}\\
PSZ2\,G108.17-11.56$_{SW}$ &&1300&1300&&180&&$-1.28\pm0.02$&\\\hline
PSZ2\,G099.48+55.60$_{N}$&\multirow{2}{*}{0.1051}&1711& 1366&\multirow{2}{*}{2405}&1.7&\multirow{2}{*}{2.81$\pm0.22$}&$-$&\multirow{2}{*}{Bo22}\\
PSZ2\,G099.48+55.60$_{S}$  &&1090&1068&&1.1&&$-$&\\\hline
PSZ2\,G113.91-37.01$_{N}$ &\multirow{2}{*}{0.3712}&1172&1210&\multirow{2}{*}{2605}&47&\multirow{2}{*}{7.58$^{+0.54}_{-0.56}$}&$-$&\multirow{2}{*}{Bo22}\\
PSZ2\,G113.91-37.01$_{S}$ &&1481&1464&&12&&$-$&\\\hline
{\bf PSZ2\,G181.06+48.47$_{NE}$} &\multirow{2}{*}{0.234}&1394&1350&\multirow{2}{*}{2694}&5.7&\multirow{2}{*}{$ 2.32^{+0.29}_{-0.25}$}&$-0.92\pm0.04$&\multirow{2}{*}{This Work}\\
{\bf PSZ2\,G181.06+48.47$_{SW}$} &&1200&1118&&5.6&&$-1.09\pm0.05$&\\\hline
PSZ2\,G205.90+73.76$_{N}$ &\multirow{2}{*}{0.4474}&664&1490&\multirow{2}{*}{2958}&2.3&\multirow{2}{*}{7.39$^{+0.55}_{-0.57}$}&$-$&\multirow{2}{*}{Bo22}\\
PSZ2\,G205.90+73.76$_{S}$ &&655&1540&&3.3&&$-$&\\\hline
PSZ2\,G233.68+36.14$_{N}$ &\multirow{2}{*}{0.3568}&570&1000&\multirow{2}{*}{2010}&17.8&\multirow{2}{*}{5.48$^{+0.65}_{-0.68}$}&$-1.31\pm0.12$&\multirow{2}{*}{Gh21}\\
PSZ2\,G233.68+36.14$_{SE}$ &&440&1000&&11&&$-0.97\pm0.13$&&\\\hline
PSZ2\,G200.95-28.16$_{NE}$ &\multirow{2}{*}{0.22}&1190&600&\multirow{2}{*}{1500}&2.31& \multirow{2}{*}{5.30$^{+0.51}_{-0.55}$}&$-$&\multirow{2}{*}{Ka17, Kn22}\\
PSZ2\,G200.95-28.16$_{SW}$ &&1020&900&&39&&$-1.21\pm0.15$&&\\\hline
PSZ2\,G286.98+32.90$_{NW}$ &\multirow{2}{*}{0.3917}&2482&600&\multirow{2}{*}{3135}&296&\multirow{2}{*}{13.69$^{+0.39}_{-0.42}$}&$-1.19\pm0.03$&\multirow{2}{*}{Bo14}\\
PSZ2\,G286.98+32.90$_{SE}$ &&1570&2700&&44.8&&$-1.36\pm0.04$&&\\\hline
PSZ2\,G277.93+12.34$_{NE}$ &\multirow{2}{*}{0.158}&660&1300 &\multirow{2}{*}{2610}&10&\multirow{2}{*}{3.6$\pm0.6$$^!$}&$-$&\multirow{2}{*}{Ko24}\\
PSZ2\,G277.93+12.34$_{SW}$ &&1640& 1300&&7.6&&$-$&\\\hline
A\,746$_{NW}$ &\multirow{2}{*}{0.214}&1800&1462 &\multirow{2}{*}{1969}&39&\multirow{2}{*}{5.34$^{+0.40}_{-0.41}$}&$-1.26\pm0.04$&\multirow{2}{*}{Ra24}\\
A\,746$_{SE}$ &&900& 492&&1.7&&$-$&\\\hline
A\,2108$_{NE}$ &\multirow{2}{*}{0.0916}&610&438 &\multirow{2}{*}{2000}&1.1&\multirow{2}{*}{3.25$^{+0.27}_{-0.28}$}&$-$&\multirow{2}{*}{Sc22, Ch24}\\
A\,2108$_{SW}$ &&300& 550&&0.5&&$-$&\\\hline
PSZ2\,G008.31-64.74$_{SE}$ &\multirow{2}{*}{0.312}&1000&2000 &\multirow{2}{*}{3200}&38.9&\multirow{2}{*}{7.75$^{+0.42}_{-0.40}$}&$-$&\multirow{2}{*}{Du24}\\
PSZ2\,G008.31-64.74$_{NW}$ &&1300& 1200&&38.9&&$-$&\\\hline
\end{longtable}
{{\bf Notes.} Column 1: Name of the cluster, Col. 2: Redshift, Col. 3: Largest linear size, Col. 4: Distance between the relic and the cluster center (X-ray peak), Col. 6: Distance between relics, Col. 6: Radio power at 1.4\,GHz, Col. 7: Mass of the cluster (\planck SZ reported masses, except where noted, Col. 8: Spectral index, Col. 9: References for radio properties. vW09: \citet{vanWeeren2009a}; vW10: \citet{vanWeeren2010}; vW11a: \citet{vanWeeren2011a}; vW11b: \citet{vanWeeren2011b}; vW12: \citet{vanWeeren2012b}; Bo09: \citet{Bonafede2009b}; Bo12: \citet{Bonafede2012}; Bo14: \citet{Bonafede2014}; Ve07: \citet{Venturi2007}; Ka12: \citet{Kale2012}; dG15: \citet{Gasperin2015}; Ka17: \citet{Kale2017}; Ro97: \citet{Rottgering1997}; St13: \citet{Stroe2013}; Bo22: \citet{Botteon2022a}; St22: \citet{Stuardi2022}; Li14: \citet{Lindner2014}; Zh20: \citet{XZhang2020}; Ho21: \citet{Hoang2021}; dG14: \citet{deGasperin2014}; Wh15: \citet{White2015}; Mi20: \citet{Mijin2020}; Du21: \citet{Duchesne2021}; Du24: \citet{Duchesne2024}; Sc22: \citet{Schellenberger2022}; Ch24: \citet{Chatterjee2024}; Ra24: \citet{Rajpurohit2024}; Ko24: \citet{Koribalski2024}; Kn22: \citet{Knowles2022}; Gh21: \citet{Ghirardini2021};  Jo21: \citet{Jones2021}; Le22: \citet{Lee2022}; Sa24: \citet{Santra2024}; St19: \citet{Stuardi2019}; dG22: \citet{deGasperin2022};  Ge17: \citet{George2017}; Ho18: \citet{Hoang2018}; Ch23: \citet{Chibueze2023}; Ho19: \citet{Hoang2019a}; $\dagger$Mass estimated from global temperature from \citet{Urdampilleta2018} with the \citet{Lovisari2020} scaling relation; $\ddagger$Mass estimated by scaling the WL $M_{200}$ from \citet{Jee2015} by a factor of 1.5; $^{\dagger\dagger}$Unscaled WL $M_{200}$ from \citet{Lee2022}; $^*$Mass estimated from global temperature from \citet{XZhang2020} with the \citet{Lovisari2020} scaling relation. $^!$Mass estimated from \planck SZ signal by \citet{Koribalski2024}.
}

\end{document}